\documentclass[runningheads]{llncs}
\usepackage[a4paper, margin=1in]{geometry} 
\usepackage{graphicx}
\usepackage[T1]{fontenc}
\usepackage{mathtools}
\usepackage{tikz}
\usepackage{amsmath}
\usepackage{filecontents}
\usepackage{amsmath,amsfonts}
\usepackage{algorithmic}
\usepackage{graphicx}
\usepackage{textcomp}
\usepackage{soul}
\usepackage{xcolor}
\usepackage{float}
\usepackage[vlined,boxed,commentsnumbered, ruled, linesnumbered]{algorithm2e}
\usepackage{subfig}
\usepackage{tikz}
\usepackage{booktabs} 
\usepackage{tablefootnote}
\usepackage{lipsum}
\usepackage{mdframed}
\usepackage{tcolorbox}
\usepackage[bottom]{footmisc}
\makeatletter
\newcommand{\removelatexerror}{\let\@latex@error\@gobble}
\newtheorem{mydef}{Definition}
\makeatother
\AtBeginDocument{%
  \providecommand\BibTeX{{%
    \normalfont B\kern-0.5em{\scshape i\kern-0.25em b}\kern-0.8em\TeX}}}
\begin{document}

\title{OptimShare: A Unified Framework for Privacy Preserving Data Sharing – Towards the Practical Utility of Data with Privacy}

\titlerunning{OptimShare: A Unified Framework for Privacy Preserving Data Sharing}

\author{M.A.P. Chamikara\inst{1,2}\orcidID{0000-0002-4286-3774} \and
Seung Ick Jang \inst{1,2} \and
Ian Oppermann\inst{3} \and
Dongxi Liu\inst{1,2} \and
Musotto Roberto\inst{2} \and
Sushmita Ruj\inst{4} \and
Arindam Pal\inst{4} \and
Meisam Mohammady\inst{5} \and
Seyit Camtepe\inst{1,2} \and
Sylvia Young\inst{6} \and
Chris Dorrian\inst{6} \and
Nasir David\inst{6}
}

\authorrunning{M.A.P. Chamikara et al.}

 \institute{CSIRO's Data61, Australia \and
Cyber Security Cooperative Research Centre (CSCRC), Australia \and
Customer, Delivery and Transformation, Department of 
  Customer Service, New South Wales, Australia \and
  University of New South Wales, Sydney, Australia \and
  Iowa State University of Science and Technology, 
  Iowa, USA \and
  Department of Health, Western Australia, Australia
}
\maketitle              
\begin{abstract}
\sloppy Tabular data sharing serves as a common method for data exchange.  However, sharing sensitive information without adequate privacy protection can compromise individual privacy. Thus, ensuring privacy-preserving data sharing is crucial. Differential privacy (DP) is regarded as the gold standard in data privacy. Despite this, current DP methods tend to generate privacy-preserving tabular datasets that often suffer from limited practical utility due to heavy perturbation and disregard for the tables' utility dynamics. Besides, there has not been much research on selective attribute release, particularly in the context of controlled partially perturbed data sharing. This has significant implications for scenarios such as cross-agency data sharing in real-world situations. We introduce OptimShare: a utility-focused, multi-criteria solution designed to perturb input datasets selectively optimized for specific real-world applications. OptimShare combines the principles of differential privacy, fuzzy logic, and probability theory to establish an integrated tool for privacy-preserving data sharing. Empirical assessments confirm that OptimShare successfully strikes a balance between better data utility and robust privacy, effectively serving various real-world problem scenarios.

\end{abstract}
\section{Introduction}

\sloppy Sharing data containing personally identifiable information (PII) may result in the exposure of sensitive personal information, thereby posing potential risks to user privacy. Data privacy, while possessing various definitions, can be characterized as ``Controlled Information Release'' in the context of data sharing and analysis~\cite{bertino2008survey}. The literature reveals several methods to ensure privacy in data sharing and analytics via ``Controlled Information Release''. Among these, disclosure control has gained prominence due to its practicality~\cite{chamikara2020privacy,mahawaga2022local}. This process entails applying various privacy preservation techniques to data prior to its release for analysis. Differential privacy (DP) is the gold standard for disclosure control mechanisms, attributed to its stringent privacy guarantees. An algorithm $M$ adheres to differential privacy if, for every pair of neighboring datasets $x$ and $y$, and all potential outputs $S$, the inequality $Pr[M(x) \in S] \leq \exp(\varepsilon) Pr[M(y) \in S] + \delta$ holds. In this context, $\varepsilon$ represents the privacy budget, indicating the privacy leak, while $\delta$ signifies the probability of model failure.

In the realm of data sharing, tabular data sharing (non-interactive data sharing) is particularly significant, as tabular data are often exchanged among agencies or released publicly in tabular format. Non-interactive data sharing poses a significant challenge due to the high degree of randomization required to maintain privacy (acceptable $\varepsilon$ values), which can result in reduced utility in the shared data~\cite{bindschaedler2017plausible}. Despite its complexity, non-interactive data sharing is crucial for enabling various opportunities, as it allows analysts to access the entire dataset for analysis without being limited to a single query output (e.g., mean). Several differentially private (DP) approaches for non-interactive data sharing, have been proposed~\cite{blum2013learning,day2015differentially,jordon2018pate,li2014differentially}. However, selecting the optimal DP approach for differentially private non-interactive data sharing is challenging due to factors such as the diversity of input datasets (e.g., statistical properties, dimensions) and the variety of applications (e.g., data clustering, deep learning)~\cite{tao2021benchmarking}. Furthermore, unanticipated data leaks may occur when privacy constraints ($\varepsilon$ and $\delta$) are relaxed to achieve higher utility~\cite{jayaraman2019evaluating}.

Prior solutions primarily emphasize one-to-one mapping between input dataset properties (e.g., table size) and output datasets, assuming fully perturbed data can deliver sufficient utility, often diverging from real-world needs~\cite{tao2021benchmarking}. However, factors such as trustworthiness levels of third parties (e.g., fully-trusted $\rightarrow$ fully-untrusted) and unique utility dynamics for diverse applications must be considered. Thus, investigating a partial data perturbation approach, where specific columns remain non-perturbed, is crucial. Differential privacy (DP) in non-interactive data sharing with a subset of the dataset (strategically chosen attributes) being released for mandated purposes has not been thoroughly explored. This is paramount in real-world contexts, such as cross-agency data sharing settings. Incorporating a non-perturbed vertical partition in the final dataset would enhance utility for custom query-based applications, but necessitates in-depth analysis concerning linkability and attack resilience, a problem we refer to as controlled partially perturbed non-interactive data sharing (CPNDS). A framework enabling CPNDS in an application-specific utility and privacy-preserving manner is indispensable. CPNDS challenges involve (1) the presence of various complex input data dynamics (e.g., categorical / non-categorical), (2) utility maintenance for diverse application demands, and (3) striking an appropriate privacy-utility balance. A unified framework-based solution addressing these concerns is required for CPNDS, but currently, no such comprehensive solutions exist.  

In addressing this issue, we present a unified multi-criterion framework-based solution, called OptimShare, to generate a practical privacy-preserving instance of an input dataset under CPNDS. We presume OptimShare operates under a central authority (a data custodian such as a government agency, hospital, or bank) with full ownership  and control over the datasets before releasing a privacy-preserving version, which is a primary requirement for CPNDS. OptimShare employs an iterative method to identify the optimal perturbed instance for release in data analytics. The empirical results demonstrate that OptimShare effectively balances utility and privacy for the selected dataset intended for release. Additionally, a comprehensive tool, available in both web-based and stand-alone versions, was developed to automate the entire CPNDS process.

\section{Background}
This section briefly discusses the background of methods utilized in OptimShare. These approaches include differential privacy and fuzzy logic.

\subsection{Data Perturbation and Differential Privacy}
OptimShare enforces data privacy through perturbation techniques, which can be classified into interactive and non-interactive approaches. Interactive approaches involve aggregated data release~\cite{dwork2006differential}, while non-interactive methods enable the release of a perturbed, privacy-preserving version of an input dataset, such as additive perturbation~\cite{dwork2014algorithmic,muralidhar1999general}, data swapping~\cite{hasan2016effective}, Privsyn~\cite{zhang2021privsyn}, PrivatePGM~\cite{mckenna2019graphical}, and DP-WGAN~\cite{xie2018differentially}. OptimShare focuses on privacy-preserving tabular data release and employs non-interactive perturbation techniques.

\subsubsection{Differential privacy}
OptimShare's objective is to enforce differential privacy (DP) on output data. DP is the most widely accepted privacy model~\cite{dwork2009differential}. DP mechanisms such as Privsyn~\cite{zhang2021privsyn}, PrivatePGM~\cite{mckenna2019graphical}, and DP-WGAN~\cite{xie2018differentially} have gained interest, with this paper focusing on PrivatePGM and DP-WGAN for tabular data generation in OptimShare.

DP-WGAN, a DP data generation technique, uses the Generative Adversarial Network (GAN) framework and the DP-SGD algorithm~\cite{abadi2016deep} to sanitize discriminator gradients during training~\cite{ganev2022robin,xie2018differentially}. PrivatePGM ~\cite{mckenna2019graphical} is a solution for privacy-preserving probabilistic graphical models (PGMs). PrivatePGM leverages differentially private algorithms to enable the analysis of sensitive data without sacrificing privacy.

Conventionally, DP uses two parameters, $\varepsilon$ (the privacy budget) and $\delta$ (the model failure probability), to constraint privacy leakage~\cite{arachchige2019local}. A randomization algorithm (DP mechanism - $M$) applied to a dataset ($D$) is guided by these parameters~\cite{arachchige2019local}.
For a mechanism to satisfy ($\varepsilon$, $\delta$)-differential privacy, it must satisfy Equation \eqref{dpeq}~\cite{arachchige2019local}, where $d$, and $d'$ are datasets differing by one record.

\begin{mydef}
A randomized algorithm $M$ with domain $\mathbb{N}^{|\mathcal{X}|}$  and
range $R$: is ($\varepsilon$, $\delta$)-differentially private for  $\delta \geq 0$  if for every adjacent datasets $d$, $d'$ $\in$ $\mathbb{N}^{|\mathcal{X}|}$
 and for any subset $S \subseteq R$, 
\end{mydef}

\begin{equation}
P[M(d) \in S] \leq e^{\varepsilon}P[M(d') \in S] + \delta
\label{dpeq}
\end{equation}

\paragraph{Postprocessing invariance property of DP}
Postprocessing invariance refers to the ability of a differential privacy ($DP$) algorithm to preserve its privacy guarantee despite additional computations on its outputs. Consequently, the result of any postprocessing on an $\varepsilon-DP$ output remains $\varepsilon-DP$~\cite{bun2016concentrated}.

\subsection{Fuzzy Inference Systems}
\label{fissection}
OptimShare employs Fuzzy Logic(FL)~\cite{gupta2015new,tran2008qos} to generate potential pairs of $(\varepsilon, \delta)$ values, conforming to the pre-established privacy requirements of a dataset. FL models imprecise definitions computationally via a fuzzy inference system (FIS) with three steps: fuzzification, rule evaluation, and defuzzification. \textbf{Fuzzification} maps crisp inputs to fuzzy values; \textbf{rule evaluation} links fuzzy input memberships to an output domain using a rule base; and \textbf{defuzzification} converts aggregated output memberships to a crisp value using methods such as the center of gravity (Equation \ref{cogequation})~\cite{gupta2015new,tran2008qos}.

\begin{equation}
COG=\frac{\int_{min}^{max}  \mu_{x} x dx}{\int_{min}^{max} \mu_{x} dx}
\label{cogequation}
\end{equation}

\section{The proposed work: OptimShare}
\label{methodology}
\begin{figure}[H]
\vspace{0cm}
\centering
\includegraphics[scale=0.7]{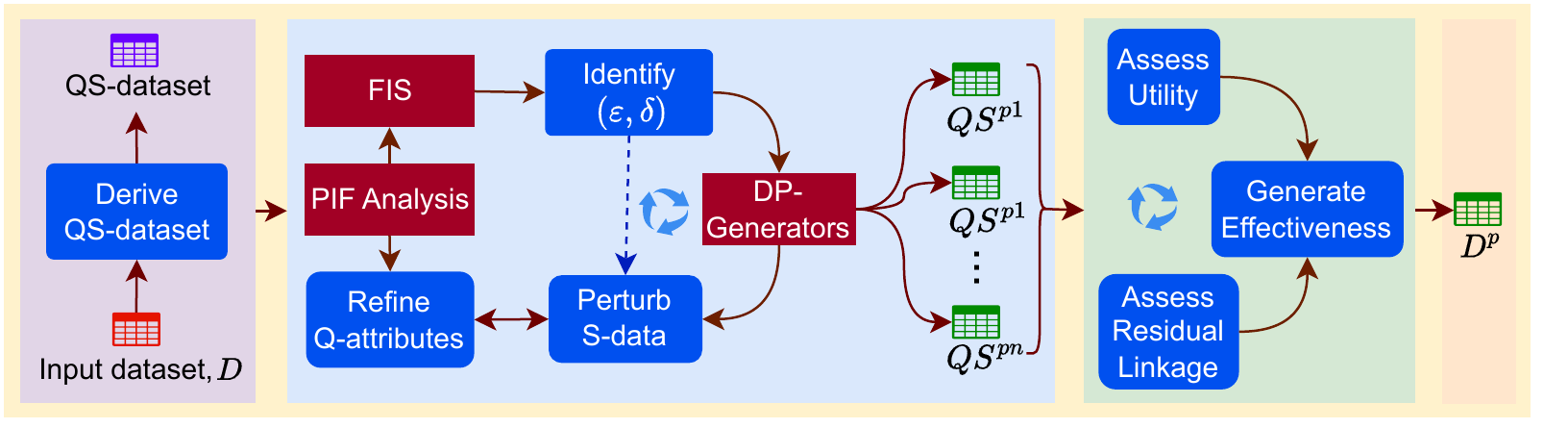}
\caption{The modular arrangement of the OptimShare framework. $D^p$ represents the perturbed output dataset of the input dataset, $D$. FIS represents the Fuzzy Inference System. PIF represents the Personal Information Factor. $QS^{pi}$ represents the intermediate perturbed instances of the $QS$ dataset.}
\label{optimshareframework}
\end{figure}

OptimShare is controlled by a central authority (a data custodian), such as a government agency, to handle Controlled Partially Perturbed Non-Interactive Data Sharing (CPNDS) with differential privacy, as illustrated in Figure \ref{optimshareframework}. The objective is to create a privacy-preserving version of the existing dataset for third-party analytics utilization. For enhanced dataset security, user role management is integrated to regulate access levels.  The focus of this paper is on the OptimShare central algorithm, presuming that the data curator has unrestricted access to the dataset and OptimShare for producing a privacy-preserving dataset.

\subsection{Problem Definition}
Given a dataset $D$ with $n$ tuples, $m$ attributes, and $r(<m)$ sensitive attributes forming $S-dataset$, $D_r$, the remaining $(m-r)$ attributes form $D_{(m-r)}$. Applying differentially private algorithm $M$ to $D_r$ generates perturbed dataset $D_r^p$ with $n$ tuples and $r$ attributes, privacy constrained by the privacy parameters of $M$. The composition of $D_r^p$ and $D_{(m-r)}$ is released as $D^p$.

\subsection{OptimShare Algorithm}

Applying perturbation directly to an input dataset $D$ using a mechanism $M$ to create a privacy-preserving dataset, $D^p$, leaves certain questions unresolved. Algorithm \ref{ppaasalgo} demonstrates the approach employed by OptimShare for generating privacy-preserving (perturbed) datasets, effectively addressing these concerns. 

\begin{center}
\scalebox{0.8}{
\begin{algorithm}[H]
\caption{OptimShare algorithm for generating a privacy-preserving dataset}\label{ppaasalgo}
\KwIn{
\begin{tabular}{ l c l } 
        $D               $ & $\gets $ & input dataset\\
		$T_{\varepsilon}$ & $\gets$ & threshold $\varepsilon$ \\
		$TN_{\varepsilon,\delta}$ & $\gets$ & total ($\varepsilon,\delta$)  \\ & & combinations\\
        $TS               $ & $\gets $ & total number of \\ & & searches\\
		$t               $ & $\gets $ & perturbed instances \\ & & per combination\\
		$A               $ & $\gets $ & application\\
		$C$ & $\gets$ & effectiveness \\ & & coefficient \\ 
		$E^T               $ & $\gets $ & effectiveness \\ & & threshold\\
\end{tabular}
}

\KwOut{
\begin{tabular}{ l c l } 
		$D^p $ & $ \gets $ & perturbed dataset \\ & & of $D$ \\
\end{tabular}
}
   
Identify identifiers ($ID$) and  quasi-identifiers ($Q$)\;
Remove $ID$ from the dataset to produce $QS$-dataset\;
Identify tuple distribution of the $QS$-dataset\;
Determine $PIF$ of $Q$-dataset\;
Determine $PIF$ of $Q$-dataset conditioned to $S$\;
Refine the $Q$ attributes and $S$ attributes\;
Generate $TN_{\varepsilon,\delta}$ combinations as $\{$($\varepsilon_1,\delta_1$),\dots,($\varepsilon_{TN},\delta_{TN}$)$\}$ (where $\varepsilon_i<\varepsilon_{i+1}$ and $\delta_i<\delta_{i+1}$)\;
\For{each ($\varepsilon_i,\delta_i$) at $TN_{\varepsilon,\delta}$ intervals in $TS$}{
  Apply DP generators ($DPA_1,\dots,DPA_n$) to generate $t$ perturbed instances\;
  Merge $Q$ to all $t$ perturbed instances ($DP_1,\dots,DP_{t}$)\;
  Generate normalized utility values of all $DP_i$\;
  Generate $t$ residual leak normalized values\;
  Find effectiveness loss value $el_i$ of each $t$ perturbed instances for $A$ using $C$ and $T_\varepsilon$\;
  Choose all $DP_i$ that satisfy  $e_i\geq E^T$ (where $e_i = 1-el_i)$\;
  }
Return $D^p$ with the highest $e_i$; 
\end{algorithm}
}
\end{center}

First, OptimShare identifies the three primary types of attributes in the input dataset, namely identifiers, quasi-identifiers, and sensitive attributes, subsequently eliminating the identifiers. Next, the distribution of tuples in the remaining dataset, referred to as the $QS$-dataset, is determined. The $QS$ attributes are then further refined. The algorithm generates combinations of privacy parameters, $\varepsilon$ and $\delta$ specific to the input dataset. Next, OptimShare employs differentially private algorithms on the sensitive portion of the dataset ($S-dataset$), leveraging each ($\varepsilon,\delta$) combination to generate perturbed instances. The effectiveness values for the perturbed $QS$-datasets are then calculated. The algorithm finally returns the perturbed dataset with the highest effectiveness value.

\subsection{The Main Steps of OptimShare Algorithm}
\label{mainsteps}
Given an input dataset $D$ with $m$ attributes and $n$ tuples, OptimShare identifies identifier attributes ($ID$) and quasi-attributes ($Q$) within $D$. To protect against direct identification, $ID$ attributes are excluded from $D$ based on their uniqueness. The dataset intended for publication after perturbation is formed by combining $Q$ and the remaining vertical partition $S$, referred to as the $QS$-dataset.

\subsubsection{Identifying initial tuple distribution of the dataset to allow $M$ to maintain the data distribution in $QS$}
\label{datadist}
The optimal clustering dynamics are found using the $k-means$ algorithm and Silhouette analysis~\cite{dinh2019estimating}, unless the input dataset is a classification dataset with existing class labels representing tuple distribution (refer to Algorithm \ref{clustdata}).

\begin{center}
    \scalebox{0.8}{
    \begin{minipage}{1.1\linewidth}
     \removelatexerror
      \begin{algorithm}[H]
\caption{Identifying original tuple distribution of the input dataset}\label{clustdata}
\linespread{1.5}
\KwIn{
\begin{tabular}{ l c l } 
		$QS$  & $\gets $ & $QS$ dataset\\
		$cn\_range$ & $\gets $ & list of cluster numbers to be searched\\
\end{tabular}
}
\KwOut{
\begin{tabular}{ l c l } 
		$T_s$  & $ \gets $ tuple status \\
\end{tabular}
}
  \For{each $cn$ $\in$ \{$cn\_range$\}} {
  run $k-means$ clustering on $QS$, where $k = cn$\;
  $s_{cn}$ = Silhouette Coefficient of $cn$\;
  }
  select the $cn$ of $maximum(s_{cn})$\; 
  return $T_s$, which is the $k-means$ cluster label of each tuple under $maximum(s_{cn})$\;
\end{algorithm}
    \end{minipage}%
}
\end{center}

\subsubsection{Identification $Q$ attributes}
\label{idqat}
A quasi-identifier ($Q$), a unique attribute set capable of distinguishing a record, could potentially facilitate linkability via auxiliary data, posing a risk to privacy leakage.

\paragraph{Declaring $Q$ attributes}
\label{qattrefine}
Selecting data-specific $Q$ attributes is challenging due to variable definitions of sensitive attributes (i.e., domain specific). OptimShare addresses this by using a global set of common $Q$ attributes ($GQ$). These attributes are then refined based on their distinguishability using the personal information factor (PIF), a measure that gauges record indistinguishability.

\paragraph{Cell surprise factor (CSF) and personal information factor (PIF)}

Ian et al. defined PIF using entropy-based KL-divergence~\cite{oppermann,oppermann2020measure}. We extend the idea and propose CSF as a bounded measure for assessing attribute impact on record indistinguishability. CSF is computed using Equations \ref{prior},  \ref{posterior}, and \ref{csf}. The CSF provides a unique method to assess how the indistinguishability of records is affected by the introduction of a specific attribute or set of attributes. PIF, bounded by [0,1], encapsulates the attribute's CSF distribution (Definition \ref{pifdefinition}).

\begin{minipage}{\textwidth}
\begin{align}
    \text{Prior}(X): &\quad \text{Prior}(X) = P(X = x) = \frac{|x|}{|X|} \label{prior} \\
    \text{Posterior}(X): &\quad \text{Posterior}(X) = P(X = x | Y = y) =  \frac{|x,y|}{|y|} \label{posterior} \\
    \text{CSF Definition}: &\quad \text{CSF} = \left | \text{Prior}(X) - \text{Posterior}(X) \right| \label{csf}
\end{align}
\vspace{0.5cm}
\end{minipage}%

Note that $CSF$ is upper bounded by $Posterior(X)$ as OptimShare only looks at the increase in indistinguishability. Hence, in all cases interested,  $Prior(X) \leq Posterior(X)$.

\begin{definition}[PIF]
\label{pifdefinition}
Let $x_i$ be the $CSF$ value bins (bounded by [0,1]) of an attribute, where $h_i$ is the number of occurrences of each $x_i$.  

Then,

\begin{equation}
PIF=\frac{\sum_{i=1}^{n} x_{i} h_{i}}{\sum_{i=1}^{n} h_{i}}
\end{equation}
\end{definition}

\subsubsection{Application of perturbation on the $QS$-dataset}
\label{qspert}
The perturbation of the $QS$-dataset is a four-step process: (1) Conduct the PIF analysis on the $Q$ attributes, (2) Refine the $Q$ and $S$ attributes per PIF outcomes, (3) Define the privacy parameters ($\varepsilon$ and $\delta$) for the $S$-dataset through the PIF analysis, and (4) Implement perturbation on $S$ data and determine the optimal perturbed instance for sharing.

\paragraph{Assessing the $Q$ attributes using PIF }
\label{assessqupif}
Calculate $QPIF_i$ (i.e., $PIF$) for all $Q$ attributes in the $Q$-dataset. Determine $QSPIF_i$ (i.e., $PIF$) for all $Q$ attributes in the $QS$-dataset to evaluate the influence of $S$ attributes on each $Q$ attribute. The difference between $QPIF_i$ and $QSPIF_i$ indicates the independence of a specific $Q$ attribute's data distribution from the $S$ attributes. The inequality $\Delta PIF_i \geq \alpha QPIF_i$ determines the extent of PIF change, where $\Delta PIF_i = QSPIF_i - QPIF_i$ and $\alpha$ is the sensitivity coefficient. If $\alpha = 1$, it means the $PIF$ leak from $Q_i$ in the $QS$ dataset is exactly $QPIF_i$, implying that $QPIF_i < 0.5$. Attributes that satisfy $\Delta PIF_i \geq QPIF_i$ are moved to the $S$-dataset for perturbation, as their distribution is significantly altered by $S$ attributes, which could otherwise risk personal information leakage.

Next, as the initial step to determine the privacy requirements of the $S$-dataset, we calculate the $PIF$ ($PIF_{Thresh}$) of the $QS$ dataset using Equation \ref{pifthresh}. In the equation, $QSMaxPIF$ is the maximum $PIF$ value returned by the $QS$ dataset.

\begin{equation}
    PIF_{Thresh} =\left\{\begin{array}{ll}
QSMaxPIF & if \  QSMaxPIF < 1 \\
1 & \text{otherwise}
\end{array}\right.  
\label{pifthresh}
\end{equation}

\paragraph{Developing a link between $PIF$ and $(\varepsilon, \delta)$}

A link between $PIF$ and $(\varepsilon, \delta)$ in terms of enforcing differential privacy can be modeled as follows:

The definition of $(\varepsilon, \delta)$-differential privacy characterizes the probabilistic bounds for a randomized algorithm or statistical mechanism $M$. For every pair of neighboring datasets $d$ and $d'$ (that differ by a single individual's data) and for every possible subset of the output space $S \subseteq Range(M)$, this model ensures that:

\begin{equation}
P[M(d) \in S] \leq e^{\varepsilon}P[M(d') \in S] + \delta
\end{equation}

where $P[M(d) \in S]$ denotes the probability that the mechanism $M$ produces an output in set $S$ with input dataset $d$.

Here, $\varepsilon$ signifies the privacy parameter (the privacy budget), and $\delta$ is a negligible quantity representing the probability of the privacy mechanism potentially violating the $\varepsilon$-privacy condition. As $\varepsilon$ approaches zero and $\delta$ is sufficiently small, a higher degree of privacy protection is conferred. Hence, we can define a privacy metric $f(\varepsilon, \delta) = (1 - \exp(-\varepsilon)) + \delta$, which serves as a suitable gauge for quantifying privacy levels. Consequently, a decrease in the value of $f(\varepsilon, \delta)$ indicates an enhanced privacy protection.

One essential property of differential privacy is its postprocessing invariance, implying that if a random mechanism $M$ guarantees $(\varepsilon, \delta)$-differential privacy, then any post-processing function $g$ applied to the output of $M$ also maintains the $(\varepsilon, \delta)$-differential privacy. Formally, if $M$ ensures $(\varepsilon, \delta)$-differential privacy, then the composed mechanism $g \circ M$ is also $(\varepsilon, \delta)$-differentially private for all functions $g$.

In the non-interactive privacy-preserving data publishing paradigm, a data curator generates a differentially private version of a dataset $D$ using a differentially private mechanism $M$. In this setting, $f(\varepsilon, \delta)$ acts as an upper bound for privacy loss, ensuring that privacy loss does not exceed $(1 - \exp(-\varepsilon)) + \delta$.

Examining a particular attribute $A \in D$, we define the ``Personal Information Factor'' ($PIF_A$) that quantifies the attribute-specific distinguishability level. For each attribute $A$, we define $\Delta_{A}$ as the increase in indistinguishability, which can be represented as:

\begin{equation}
\Delta_{A} = Posterior(A) - Prior(A)
\end{equation}

The relationship between $PIF_A$ and $\Delta_{A}$ is given by:

\begin{equation}
PIF_A  
= \frac{\sum_{i=1}^{n} \Delta_{A_i} h_{i}}{\sum_{i=1}^{n} h_{i}}
\end{equation}

where $\Delta_{A_i}$ represents the increase in indistinguishability for the attribute $A$ in the $i$-th bin with $h_i$ occurrences.

Utilizing $PIF_A$ for each attribute, we can introduce the privacy measure $f_A$ as follows:

\begin{equation}
f_A(PIF_A, \delta) = PIF_A + \delta.
\end{equation}

Consequently, we can derive a privacy measure for the entire dataset $D$ using the maximum Personal Information Factor ($PIF_{Thresh}$) over all attributes in $D$. Hence, the privacy measure for the dataset can be defined as:

\begin{equation}
f_D(\varepsilon, \delta) = PIF_{Thresh} + \delta.
\end{equation}

$f_D(\varepsilon, \delta)$ signifies an upper bound to privacy loss upon the release of the dataset and provides a quantitative control mechanism balancing data utility and privacy protection. $PIF_{Thresh} = max(PIF_{A_i})$ signifies the maximum PIF across all attributes, indicating the dataset's potential to satisfy privacy parameters without any attribute surpassing this threshold. A fuzzy model can now be utilized to represent this relationship between $PIF_{Thresh}$ and $(\varepsilon, \delta)$.

\paragraph{Determination of the privacy parameters ($\varepsilon$ and $\delta$) for $S$-dataset perturbation}
\label{detepdelta}

\begin{figure}
	\centering
	\subfloat[Fuzzy membership functions]{\includegraphics[width=0.25\textwidth, trim=0cm 0cm 0cm 0cm]{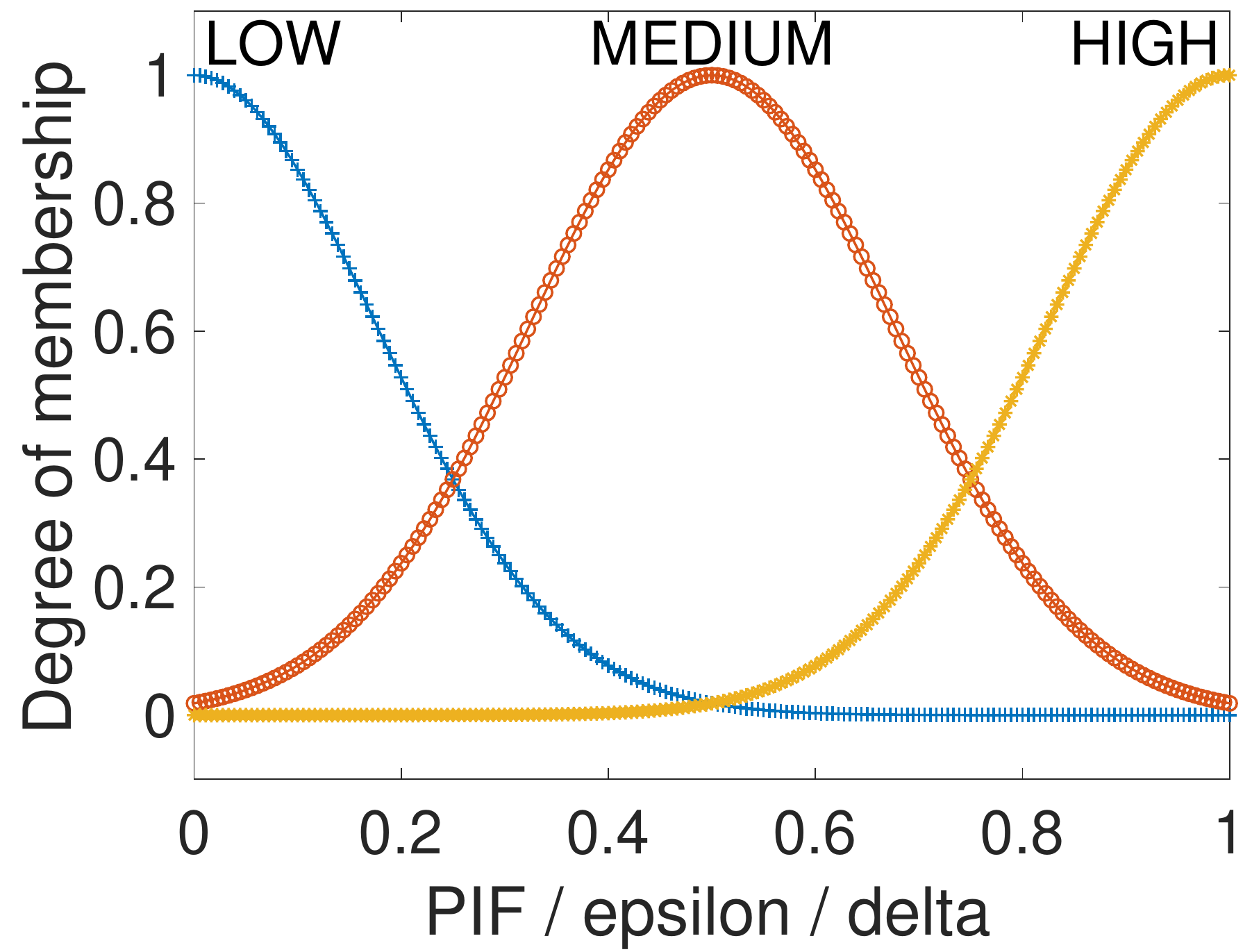}\label{fuzzyfunc}}
	\hspace{8mm}
	\subfloat[A 3D view of the rule-surface]{\includegraphics[width=0.25\textwidth, trim=0cm 0cm 0cm 0cm]{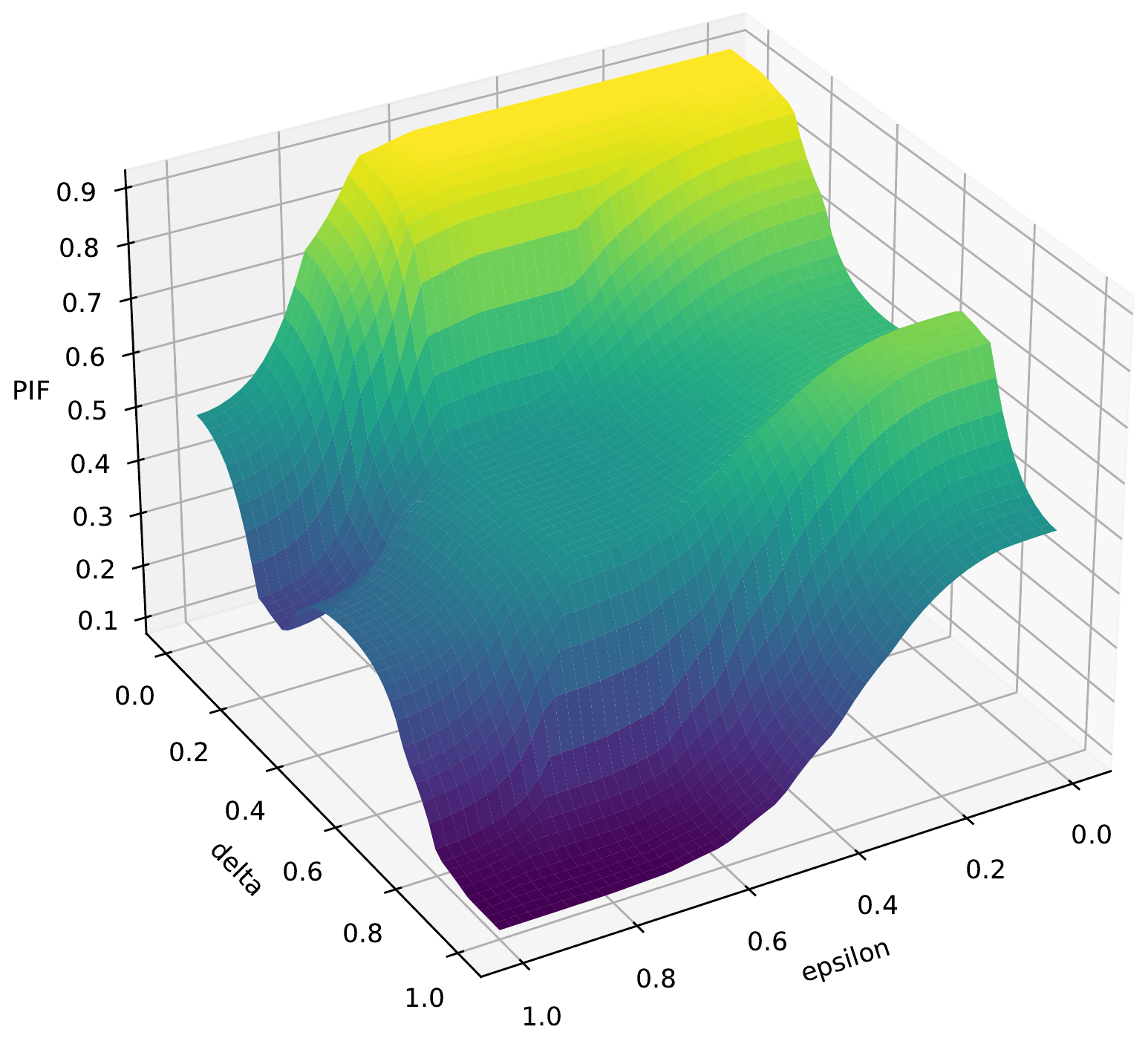}\label{fuzzysurf1}}
	\hspace{8mm}
	\subfloat[A 2D view of the rule-surface]{\includegraphics[width=0.25\textwidth, trim=0.3cm 0cm 0cm 0cm]{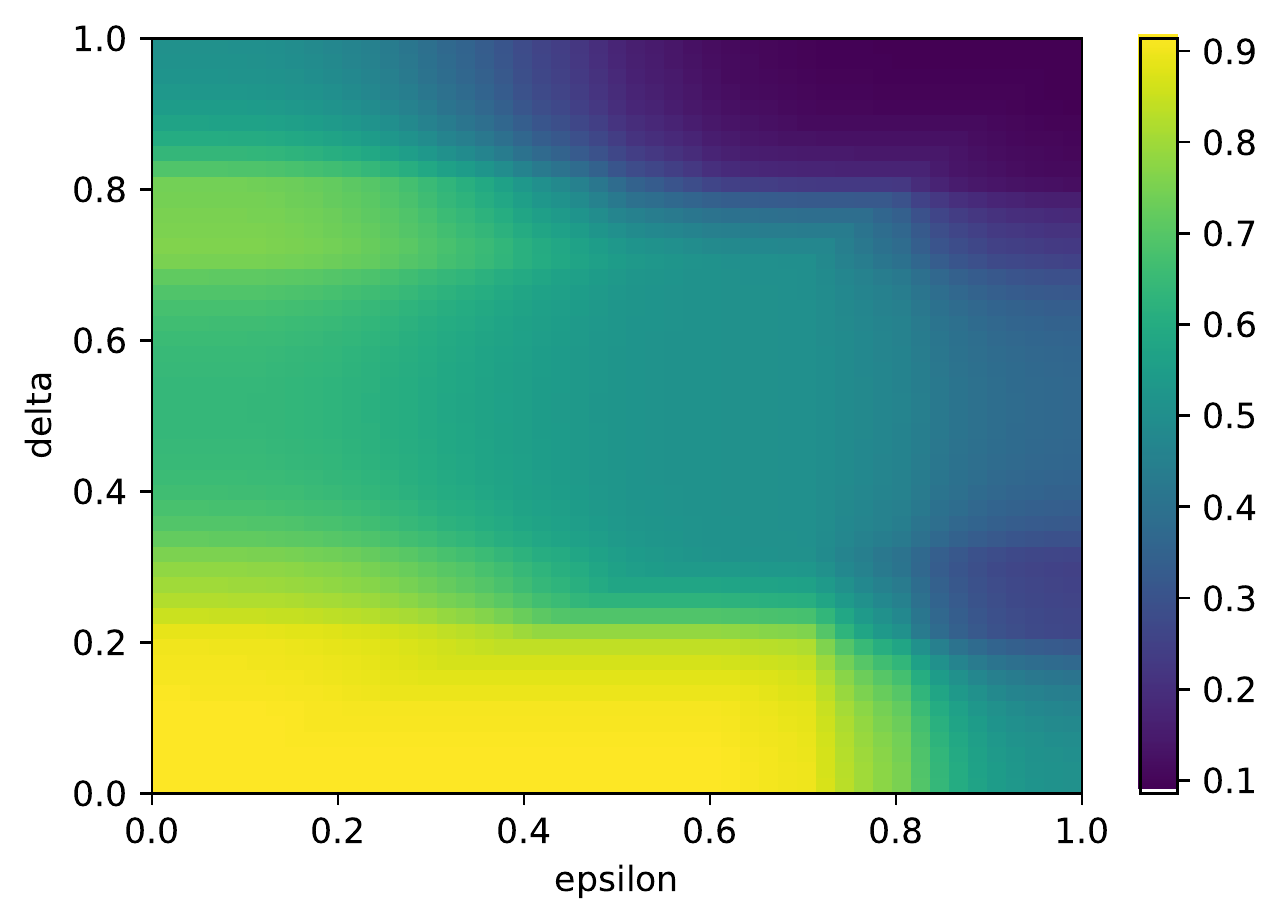}\label{fuzzysurf2}}
 	\caption{The mapping between the three fuzzy variables and the change of PIF against the changes of $\delta$ and $\varepsilon$.}
 	\label{fuzzysurf}
 	\medskip
    \small
    \vspace{-1cm}
\end{figure}

Optimshare employs a fuzzy inference system (FIS) for determining suitable $\varepsilon$ and $\delta$ inputs for the $S$-dataset from $PIF_{Thresh}$. Higher $PIF$ ($PIF_{Thresh}$) suggests enhanced distinguishability of the $QS$ dataset and, consequently, greater privacy need for the $S$ data via increased perturbation. We model an FIS to encapsulate the relationship between $PIF, \varepsilon,$ and $\delta$. Each of the three fuzzy variables have three Gaussian-shaped membership functions $(LOW, MEDIUM, HIGH)$ signifying different input value ranges and facilitating a gradual shift between functions for a broader value spectrum (see Figure \ref{fuzzyfunc}). The mean ($\mu$) and standard deviation ($\sigma$) for LOW, MEDIUM, and HIGH are respectively set as ($\mu=0,\sigma=1$), ($\mu=0.5,\sigma=1$), and ($\mu=1,\sigma=1$).

\begin{equation}
\begin{aligned}
& \textbf{Rule 1:}\ IF (\varepsilon = LOW) \ THEN\ (PIF = HIGH) \\ 
& \textbf{Rule 2:}\ IF (\delta = LOW)\  THEN\ (PIF = HIGH) \\
& \textbf{Rule 3:}\ IF (\varepsilon = MEDIUM\ AND\ \delta = MEDIUM)\  THEN\ (PIF = MEDIUM) \\
& \textbf{Rule 4:}\ IF (\varepsilon = HIGH\ )\  THEN\ (PIF = LOW) \\
& \textbf{Rule 5:}\ IF (\delta = HIGH)\  THEN\ (PIF = LOW) \\
\end{aligned}
\label{rules}
\end{equation}

Figure \ref{fuzzyfunc} depicts the fuzzification of variables $\varepsilon$, $\delta$, and $PIF$, with the y-axis quantifying their degree of membership. A fuzzy rule base, providing the foundation for fuzzy inference, is established next. Equation \ref{rules} represents the proposed FIS rules, defined by the IF-THEN convention (e.g., $IF (\varepsilon = MEDIUM\ AND\ \delta = HIGH)\ THEN\ (PIF = MEDIUM)$). The FIS rule evaluation step fuses fuzzy conclusions into one via the fuzzy rule base, applying $MAX-MIN$ ($OR$ for $MAX$ and $AND$ for $MIN$) operation. The minimum among membership levels is considered for each rule, while the maximum fuzzy value from all rule outputs determines the value conclusion.

Figure \ref{fuzzysurf} depicts the rule surface between the three fuzzy variables. As shown in the rule surface, higher values of PIF correspond to lower values for $\varepsilon$ and $\delta$. The final step of the FIS is the defuzzification based on the rule aggregated shape of the output function. We use the centroid-based technique to obtain the final defuzzified output value, where $x = output$ and $\mu_{x} =$ degree of membership of $x$. As depicted in the fuzzy-rule surface (refer to Figure \ref{fuzzysurf}), a single $PIF$ value corresponds to a collection of $(\varepsilon, \delta)$ combinations.  

\subsubsection{Application of perturbation on the $S$-dataset}
\label{dpsection}
OptimShare generates a list of ($\varepsilon$ and $\delta$) combinations matching the input dataset's $PIF_{Thresh}$. With a specific ($\varepsilon$ and $\delta$) pair, it perturbs the $S$-dataset, generating a set number of perturbed instances that reflect the data distributions (refer to Section \ref{datadist}). Each perturbed version is $min-max$ rescaled to the original attribute $min-max$ values, then merged with the $Q$-dataset to create perturbed $QS$ datasets.

\subsubsection{Privacy analysis}
\label{privindist}
Our threat model assumes the worst-case scenario, with the attacker having full knowledge of the $Q$ attributes in the perturbed $QS-dataset$, to assess residual linkage risk. We define a similarity group, $SG_k$, as a collection of identical records ($Q$) in the $QS$ dataset. We compute the cosine similarity ($CS^i_r$) between original and perturbed $S$ attributes for each record ($r_i$) in each $SG_k$. The worst-case record linkability is then defined as per Definition \ref{deflink}.

\begin{center}
\scalebox{0.9}{
\begin{tcolorbox}[ sharp corners, colback=green!20, colframe=green!80!blue, title=The Threat Model]
The adversary has a complete knowledge (e.g., record order, attribute domain) of the $Q$ attributes. This assumption leads to a worst-case linkage risk by enabling the adversary to explore the linkability of the records through $Q$ (and perturbed $S$) attributes based on the tuple similarity. The knowledge acquired will subsequently be leveraged by the adversary to extract the sensitive information of individuals.
\end{tcolorbox}
}
\end{center}

\begin{definition}[Record linkability]
\label{deflink}

Let $R$ be the set of all rows in the perturbed ($P$) and original ($D$) datasets. If $q^\alpha = q^\beta$ for some $\alpha, \beta \in R$ and $q \in Q$, take $(q^\alpha, s^\alpha) \in SG$. For each $SG_k \in SG$ compute $CS^i_k$ for some $i \in R_{SG_k}$, where $R_{SG_k}$ is all records in $SG_k$. If $CS^i_k \leq CS^j_k \ \forall \ j \in R_{SG_k}$, then $CS^i_k \in L$, where $L$ is the set of linkable records.

\end{definition}

\begin{theorem}
For any $\alpha, \beta \in R$ such that $q^\alpha = q^\beta$ for some $q \in Q$, the probability that $(q^\alpha, s^\alpha)$ and $(q^\beta, s^\beta)$ are in the same similarity group and $(q^\alpha, s^\alpha)$ is linkable is small. Refer to Section \ref{prooflinkability}, Proof 1, for the proof.
\end{theorem}

\begin{theorem}
OptimShare framework satisfies $\varepsilon$-differential privacy when the following inequality holds. Refer to Section \ref{optimsharedpproof}, Proof 2, for the proof.
\begin{center}
    \scalebox{1.2}{$
$$\frac{P[(q^\alpha, s^\alpha) \in SG \ \land \ CS^i_k \leq CS^j_k \ \forall \ j \in R_{SG_k}]}{P[(q^\alpha, s^\beta) \in SG \ \land \ CS^i_k \leq CS^j_k \ \forall \ j \in R_{SG_k}]} \leq e^\varepsilon$$
    $}
\end{center}

\label{theroemdp}
\end{theorem}

\subsubsection{Analysis of Utility and Effectiveness in Data Perturbation}

The utility can be measured based on any measurement such as accuracy, precision, recall, and ROC area ($KL$-divergence for generic scenarios) normalized within [0,1]. Consider $KL_x$ as the KL-divergence between a perturbed attribute, $x^p_i\in S$, and its unperturbed version, $x_i$. The maximum $KL_x$ is the dataset's $KL$-divergence, indicating the highest distribution difference. The utility loss $U_l$ quantifies the utility reduction resulting from data perturbation, given an original utility $U_o$ and a utility $U_p$ after the perturbation.

The effectiveness of perturbation is gauged by the normalized residual linkage leak $P_N$ and the $\varepsilon$-threshold $T_\varepsilon$ set by the OptimShare curator. The dataset is not suitable for release if $P_N$ is too high, which is calculated as $\frac{\varepsilon L}{T_{\varepsilon}}$ if $T_{\varepsilon} > \varepsilon L$, or 1 otherwise, where $L$ represents linkable records.

The effectiveness loss ($E_l$) of a perturbed dataset is defined as a weighted measure of $U_l$ and $P_N$, calculated by $E_l = C \ U_l + (1-C) \ P_N$. Here, $C$ determines the emphasis on linkage protection (high $C$) versus utility preservation (low $C$). The ranges of $E_l$ are dependent on $P_N$ and $U_l$ values: For Low $P_N$ and Low $U_l$: $E_l$ is in [0, $C$].
For High $P_N$, low $U_l$: $E_l$ is in [$C$, 1].
For Low $P_N$, high $U_l$: $E_l$ is in [$1-C$, 1]. For High $P_N$ and High $U_l$: $E_l$ is in [$C$, 1]. In our study, we set $C$ to $0.5$ to treat residual linkability leak and utility as equally crucial.

\section{Results and Discussion}
\label{results}
This section outlines the process of implementing OptimShare as a live tool (a usable product in the real world) and setting up the experiments. Additionally, we discuss the intermediate steps and dynamics of OptimShare.

\subsection{Implementation}
\label{thetool}
We developed two versions of OptimShare (using Python 3.8): a server-based for large-scale settings and a stand-alone for single-computer use. Figure \ref{serverversion} and \ref{standalone} show the screen captures of the stand-alone and the server version.

\begin{figure}
\vspace{-0cm}
\centering
\includegraphics[scale=0.4]{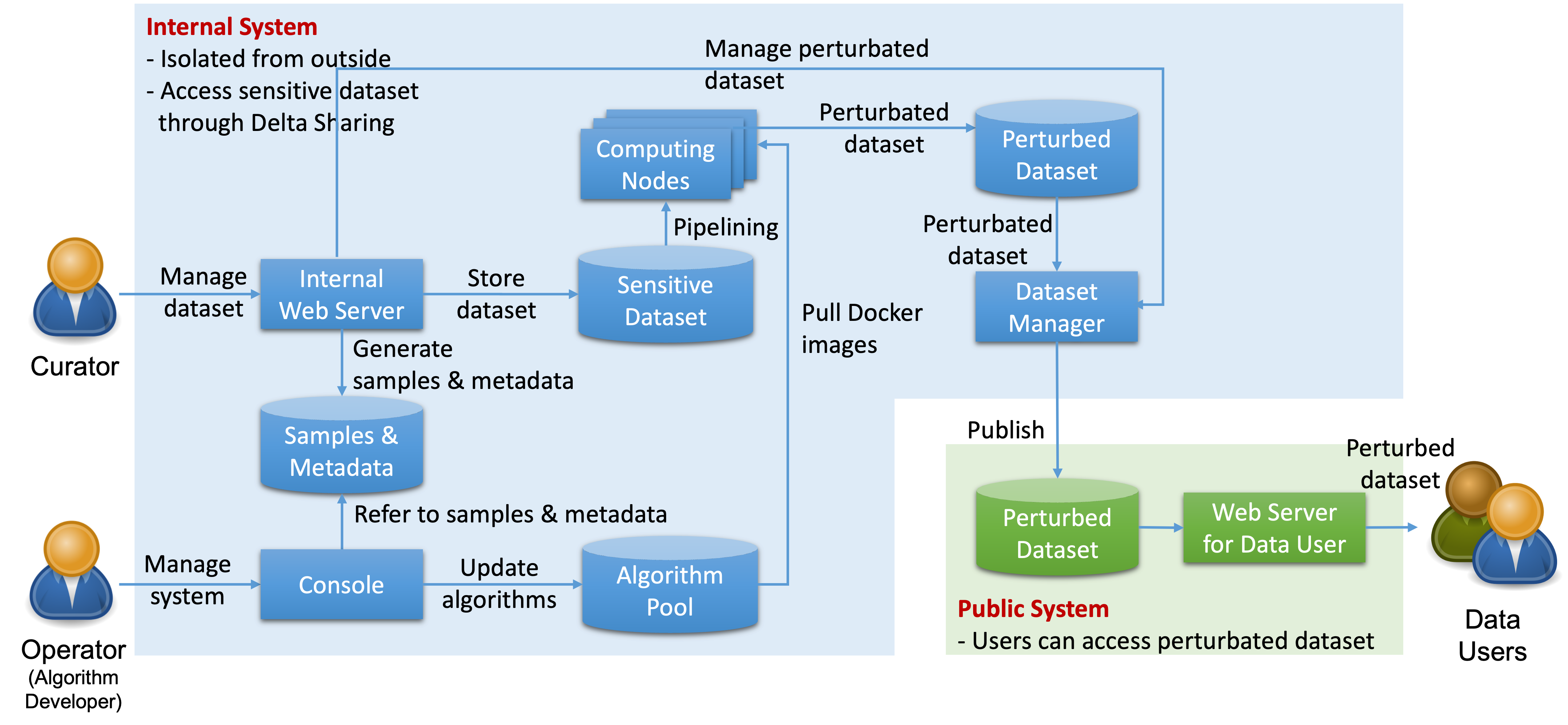}
\caption{System design of the  OptimShare server-based version}
\label{systemdesign}
\vspace{-0.5cm}
\end{figure}

\subsubsection{OptimShare server-based version}

Figure \ref{systemdesign}  outlines a server-based system design with three user roles: curator (the data custodian), operator (admin), and data user, each with distinct privileges. Curators own and manage original datasets, applying OptimShare data perturbation, auditing, and publishing perturbed datasets for data users. Operators, as administrators, manage the algorithms while being restricted from accessing the original datasets. Data users consume the perturbed datasets approved by curators. The system ensures security and data privacy by allowing dataset owners exclusive control and isolating servers from external access. OptimShare uses Docker containers to store the privacy-preserving algorithm for scalability and continuous integration and deployment (CI/CD). The dataset manager then pushes the published datasets to the public system, where data users can only access approved, perturbed datasets.

\subsection{Experiments}

This section discusses the generation of perturbed privacy-preserving datasets using the datasets and configurations mentioned in Table \ref{datasettb}. The experiments were performed on an Apple MacBook Pro with an M1 Max and 32GB of RAM, with all plots generated automatically by our live tool (see Section \ref{thetool}).

\begin{table}
\vspace{-0.5cm}
\centering
\begin{minipage}[t]{1\textwidth}
\centering
\caption{Datasets used for the experiments. Note: all the datasets are tabular.}   
\label{datasettb}
\resizebox{1\textwidth}{!}{$
\begin{tabular}{ l c c c c l }
\toprule
\bfseries Dataset & \bfseries Abbr. & \bfseries Records & \bfseries Attributes & \bfseries Classes & \bfseries Global Q Attributes \\
\midrule
NHANES diabetes Kaggle\footnotemark[1] & NHDS & 4,412 & 17 & 2 & `BPQ020', `RIAGENDR', `ALQ120Q', `LBXTC'\\    
Wine Quality\footnotemark[2] & WQDS & 4,898 & 12 & 7 & `free sulfur dioxide', `total sulfur dioxide' \\
Page Blocks Classification\footnotemark[3] & PBDS  & 5,473 & 11 & 5 & `at1', `at2', `at10' \\
Letter Recognition\footnotemark[4] & LRDS & 20,000 & 17 & 26 & `lettr', `x-box', `y-box', `width', `high', `xy2br' \\
Statlog (Shuttle)\footnotemark[5] & SSDS & 58,000 & 9 & 7 & `b', `d', `i' \\
Credit Score Kaggle\footnotemark[6] & CSDS & 150,000 & 11 & 2 & `ID', `\#ofOCLL', `\#ofT90DL', `\#RELL', `\#ofT60DPDNW', `\#ofDependents' \\
\bottomrule
\end{tabular}
$}
\end{minipage}%

\vspace{-1cm}
\end{table}
\footnotetext[1]{https://www.kaggle.com/cdc/national-health-and-nutrition-examination-survey}
\footnotetext[2]{https://archive.ics.uci.edu/ml/datasets/Wine+Quality}
\footnotetext[3]{https://archive.ics.uci.edu/ml/datasets/Page+Blocks+Classification}
\footnotetext[4]{https://archive.ics.uci.edu/ml/datasets/Letter+Recognition}
\footnotetext[5]{https://archive.ics.uci.edu/ml/datasets/Statlog+\%28Shuttle\%29}
\footnotetext[6]{https://www.kaggle.com/c/GiveMeSomeCredit/data?select=cs-training.csv}

\subsubsection{The configurations of OptimShare}
\label{OptimShareconfig}

In the experiments, the primary parameters for OptimShare were set as follows: $T_{\varepsilon}$ = 8, $P_l$ = 0.01\% ($\delta = (1 / (100 \times \text{number of rows of } D)) \times P_l$), $TN_{\varepsilon,\delta}$ = 12, $t$ = 4, $A$ = ``classification - GaussianNB'', $C$ = 0.5, $E^T$ = 0.5 (see Section \ref{methodology} for parameter details). Global $Q$ attributes used for each dataset are provided in Table \ref{datasettb}. All settings remained constant in all experiments, ensuring uniformity for unbiased results. DP-WGAN (focusing non-categorical attributes) and PrivatePGM (focusing categorical attributes) were used for $S$ data perturbation.

\begin{figure}[htbp]
\vspace{-0.5cm}

\begin{center}
  \captionsetup{type=table} 
  \begin{minipage}{0.65\textwidth}
    \centering
    \caption{Experiment results}   
    \resizebox{1\textwidth}{!}{$
    \begin{tabular}{ l c c c c c}
    \toprule
        {\bfseries Dataset} & 
        \bfseries{\# of Records} & 
        \bfseries{\begin{tabular}[c]{@{}c@{}}Total Processing\\ Time (sec)\end{tabular}} & 
        \bfseries{\begin{tabular}[c]{@{}c@{}}Record Processing\\ Time (ms)\end{tabular}} & 
        \bfseries{\begin{tabular}[c]{@{}c@{}}Average\\ Utility\end{tabular}} &
        \bfseries{\begin{tabular}[c]{@{}c@{}}Average\\ Effectiveness\end{tabular}}  \\
    \midrule
NHDS & 4,412 & 517.6 & 117.3 & 0.784 & 0.725\\    
WQDS & 4,898 & 471.1 & 96.2 & 0.801 & 0.900\\
PBDS & 5,473 & 546.6 & 99.9 & 0.717 & 0.859\\
LRDS & 20,000 & 4600.9 & 230.0 & 0.942 & 0.971\\
SSDS & 58,000 & 11,579.3 & 199.6 & 0.925 & 0.962 \\
CSDS & 150,000 & 49,867.2 & 332.4 & 0.806 & 0.903 \\
    \bottomrule
    \end{tabular}
    $}
    \label{resulttb}
  \end{minipage}%
  \captionsetup{type=figure} 
  \begin{minipage}{0.30\textwidth}
    \centering
    \includegraphics[width=\textwidth]{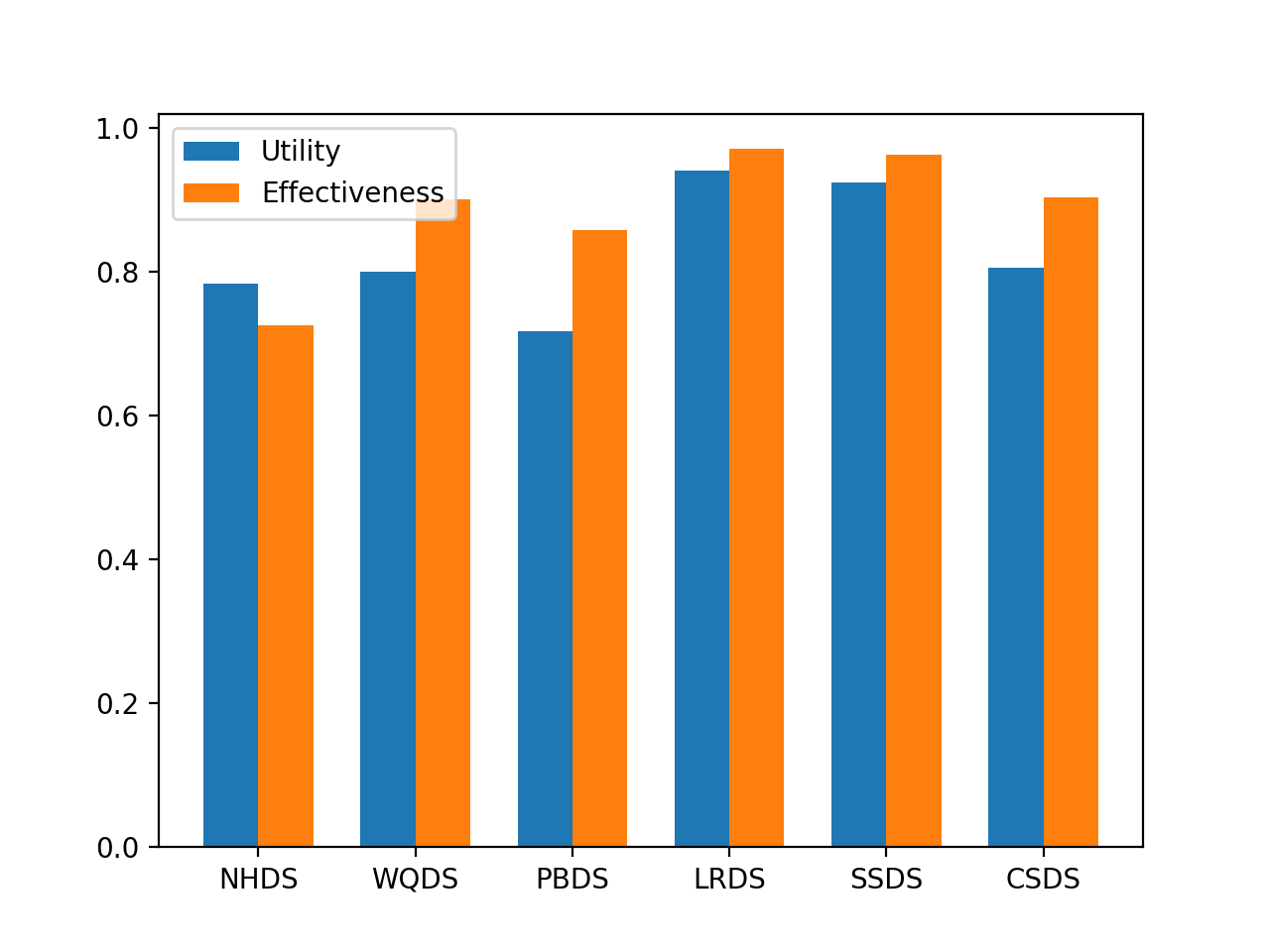}
    \caption{Average utilities and effectivenesses}
    \label{avgutileffect}
  \end{minipage}
\end{center}
\vspace{-1cm}
\end{figure}

\subsubsection{Results}
Table \ref{resulttb} displays the processing time for each dataset and the averages of utilities and effectivenesses for the privacy-preserving datasets generated by OptimShare, while Figure \ref{avgutileffect} plots the averages of utilities and effectivenesses. The datasets exhibit high effectiveness due to high utility and minimal residual data linkability. The average time complexity of the OptimShare algorithm is $O(nl)$, where $n$ is the number of records and $l$ is the product of $TS$ and $t$. The training time of DP-WGAN and PrivatePGM models increases according to the number of records. Thus, as the number of records increased, the total processing time increased accordingly.

\subsection{Dynamics of OptimShare Algorithmic Steps}
\begin{figure}
	\subfloat[CSF analysis of the dataset]{\includegraphics[width=0.22\textwidth, trim=0cm 0cm 0cm 0cm]{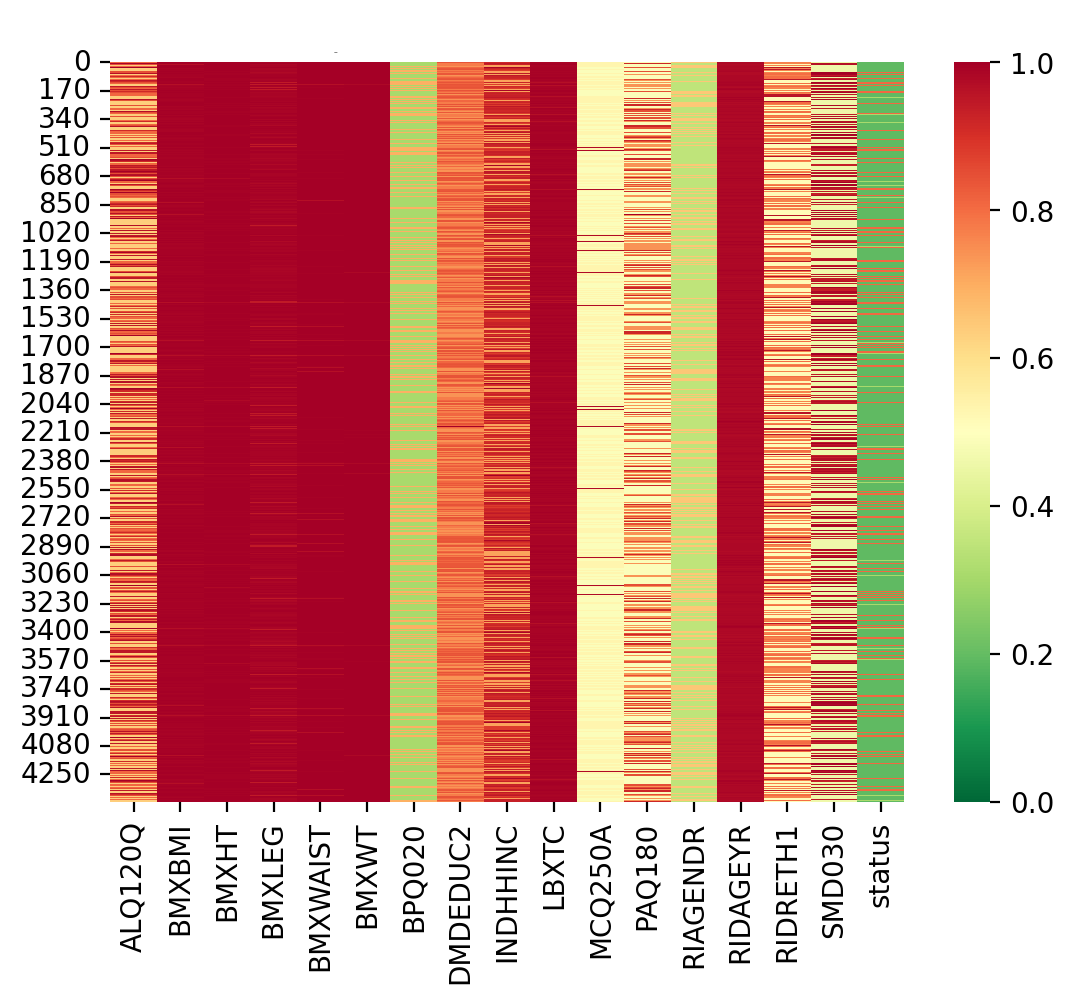}\label{csfplot}}
	\hfill
	\subfloat[PIF analysis of the dataset]{\includegraphics[width=0.22\textwidth, trim=0.3cm 0cm 0cm 0cm]{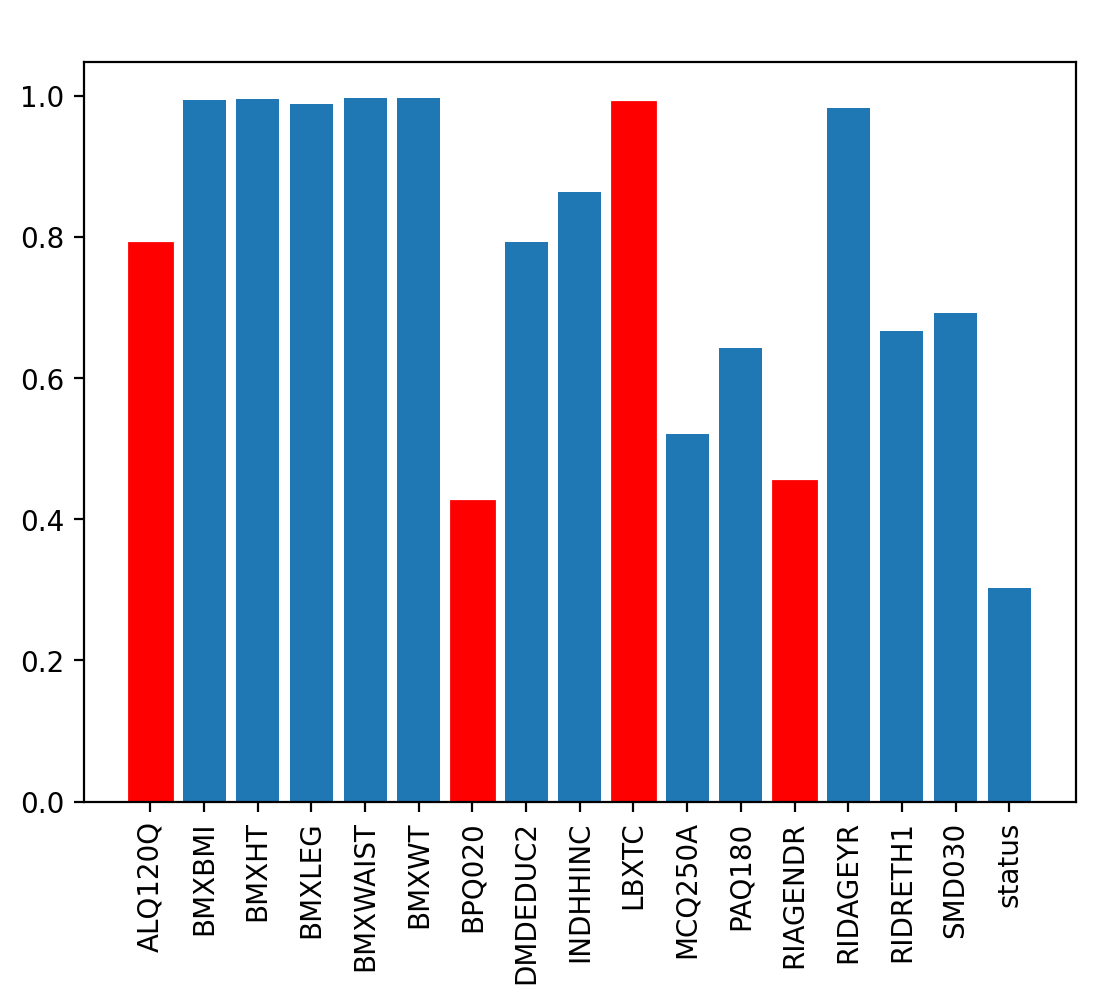}\label{pifplot}}
 	\medskip
    \small
\hfill
	\subfloat[CSF analysis of the $Q$ attributes]{\includegraphics[width=0.22\textwidth, trim=0cm 0cm 0cm 0cm]{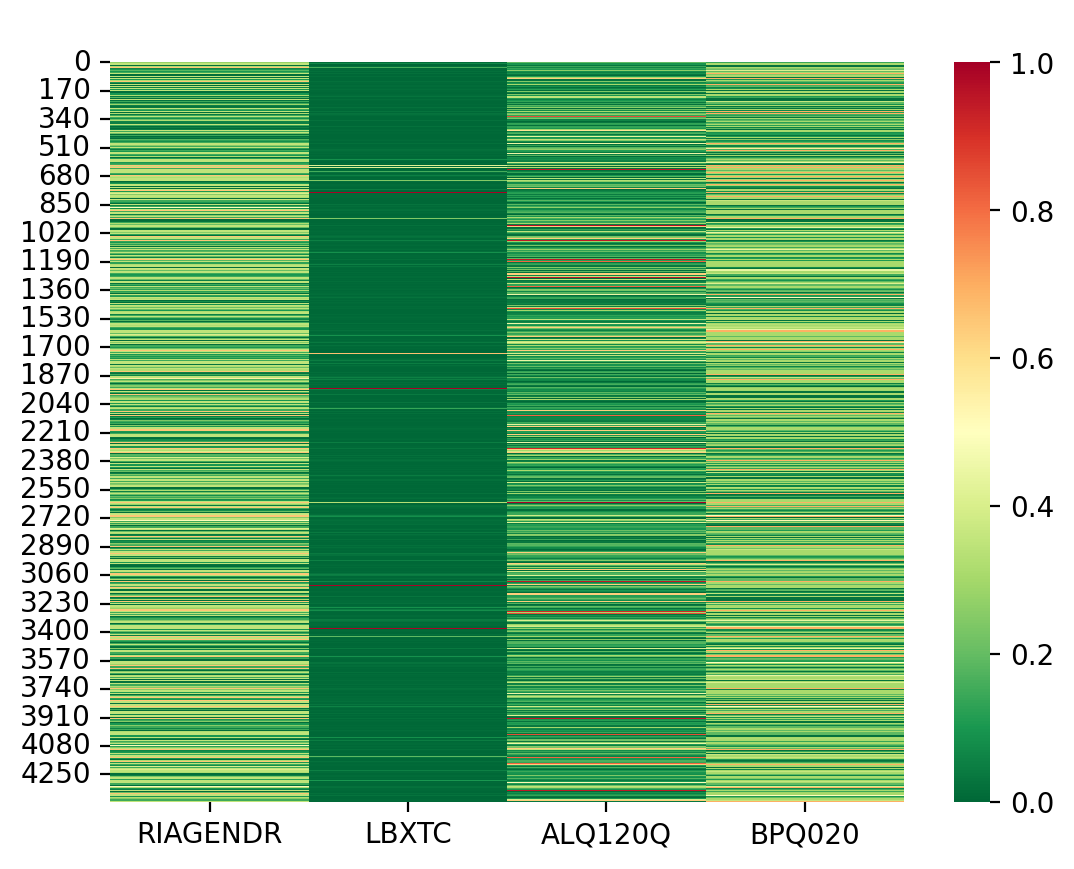}\label{csfplotqat}}
	\hfill
	\subfloat[PIF analysis of the $Q$ attributes]{\includegraphics[width=0.22\textwidth, trim=0.3cm 0cm 0cm 0cm]{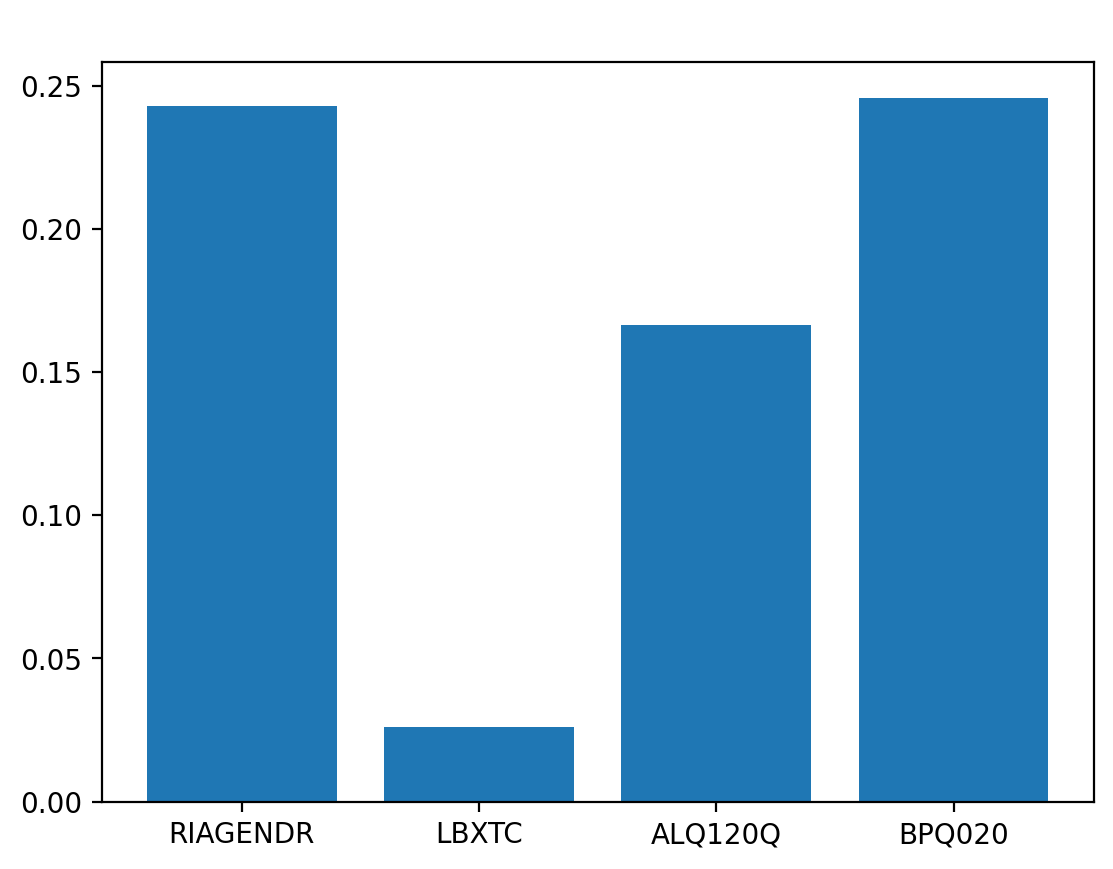}\label{pifplotqat}}
 	\caption{The CSF and PIF analysis of the input dataset and the CSF and PIF analysis of the $Q$ attributes. Note: The red bars in (b) represent the $Q$ attributes.}
 	\label{csfpifplotsqat}
 	\medskip
    \small

    \begin{minipage}[t]{0.45\textwidth}
        \centering
        \subfloat[CSF analysis of the refined set of  $Q$ attributes]{\includegraphics[width=0.45\textwidth, trim=0cm 0cm 0cm 0cm]{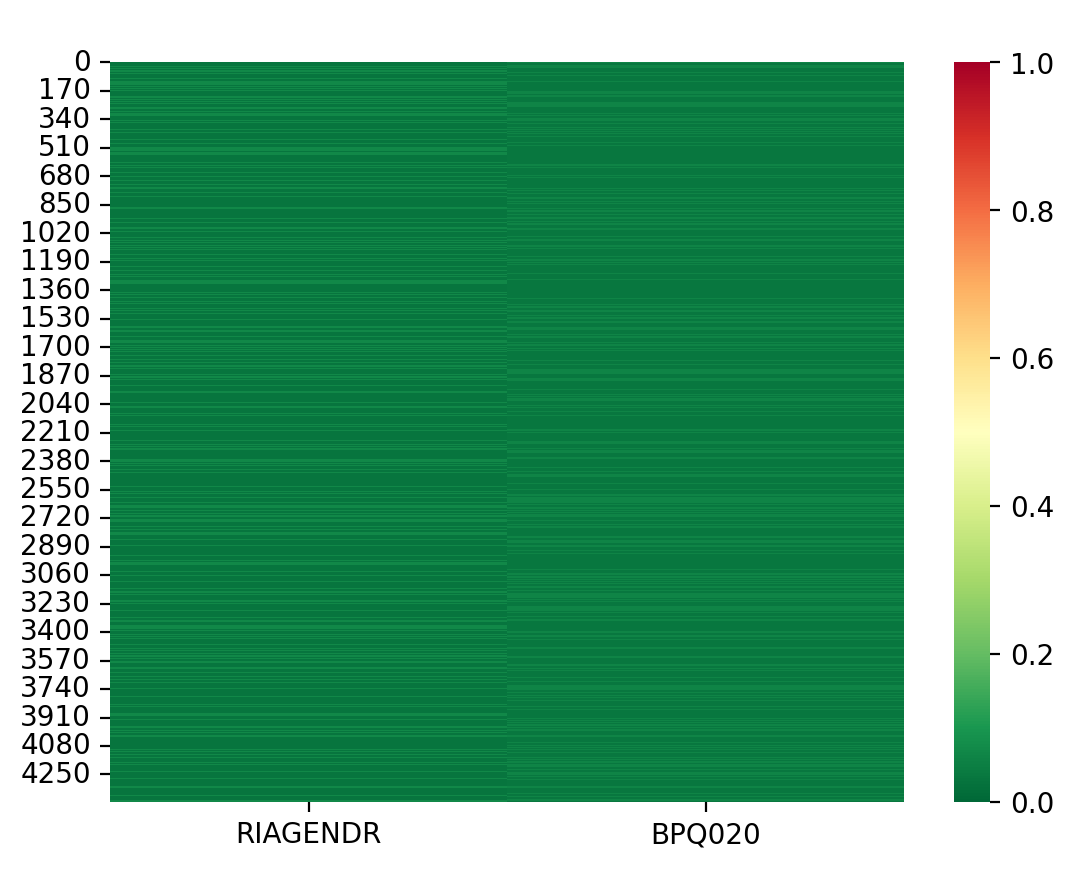}\label{refcsfplotq}}
        \hfill
        \subfloat[PIF analysis of the refined set of  $Q$ attributes]{\includegraphics[width=0.45\textwidth, trim=0cm 0cm 0cm 0cm]{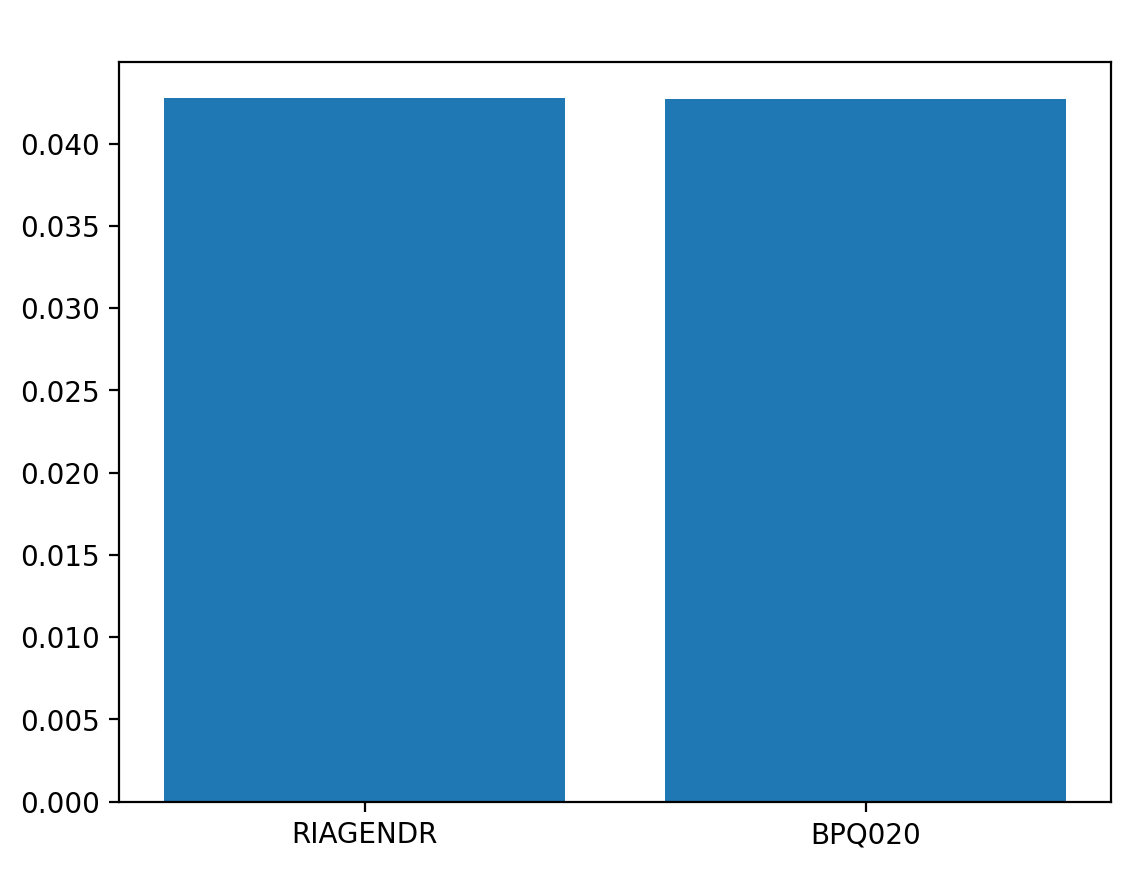}\label{refpifplotq}}
        \caption{The CSF and PIF analysis of the refined set of $Q$ attributes}
        \label{refcsfpifplotsq}
    \end{minipage}
    \hspace{0.04\textwidth}
    \begin{minipage}[t]{0.45\textwidth}
        \centering
        \subfloat[Utility analysis of privacy-preserving datasets]{\includegraphics[width=0.45\textwidth, trim=0cm 0cm 0cm 0cm]{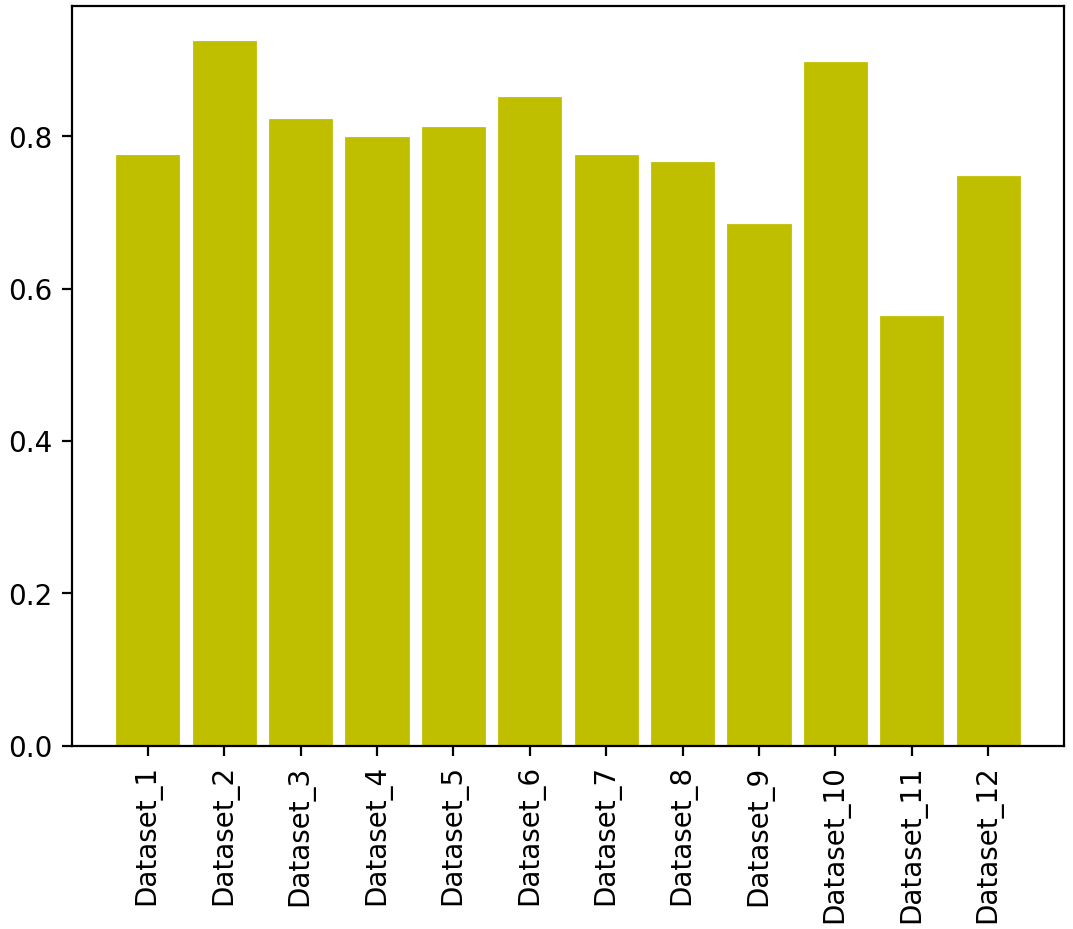}\label{utilityplot}}
        \hfill
        \subfloat[Effectiveness analysis of privacy-preserving datasets]{\includegraphics[width=0.45\textwidth, trim=0cm 0cm 0cm 0cm]{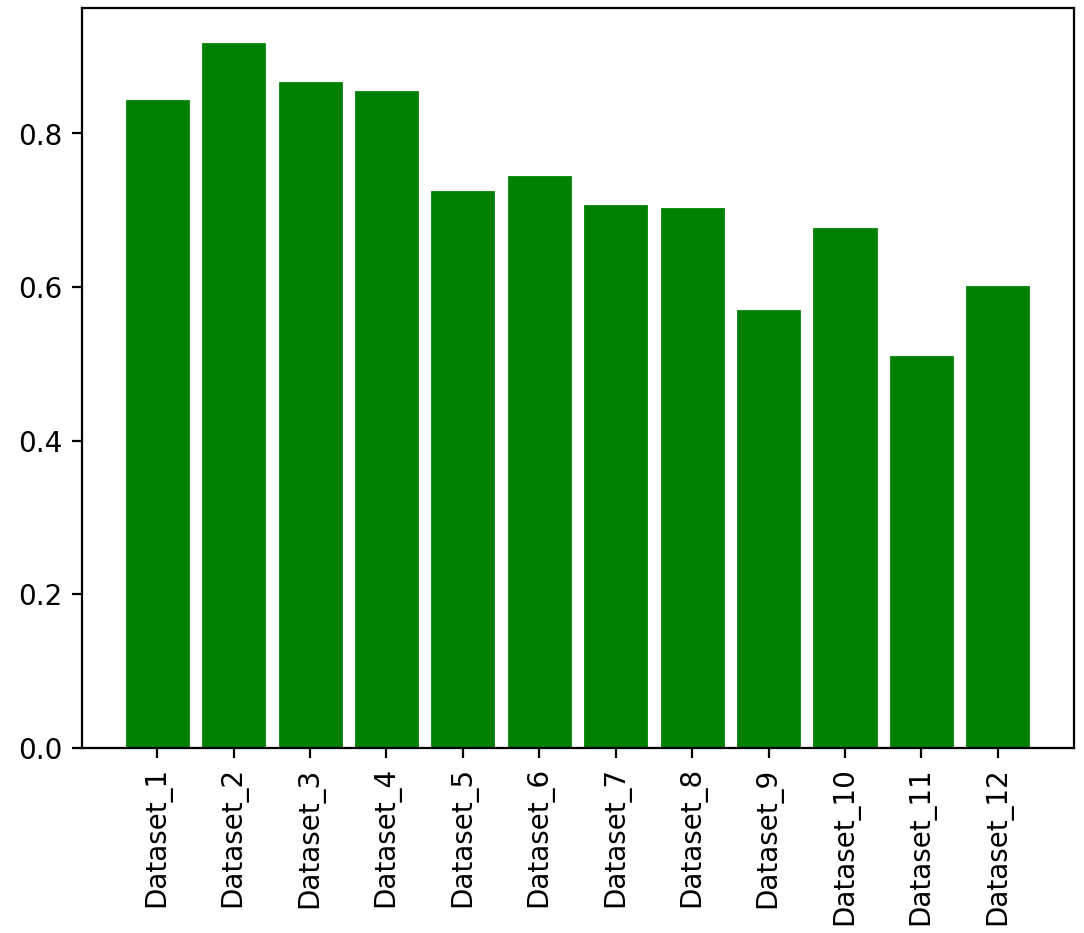}\label{effectivenessplot}}
        \caption{A comparison between utility and effectiveness of the privacy-preserving datasets generated by OptimShare}
        \label{utilityeffectiveness}
    \end{minipage}
\end{figure}

In this section, we discuss the intermediate steps involved in OptimShare's process for generating and releasing a private dataset. By understanding these steps, we can gain a comprehensive understanding of the experimental dynamics behind the process. As discussed in Section \ref{methodology}, one of the fundamental components of OptimShare is the determination of privacy requirements. This is done through $PIF$ analysis, as explained in Section \ref{methodology}. As shown in Figure \ref{csfplot}, the NHDS dataset (refer to Table \ref{datasettb})  shows extreme $CSF$ values (represented by dark red in the heatmap) in certain attributes (e.g., BMXBMI, BMXHT ), whereas certain other attributes such as BPQ020 shows lower $CSF$ values (represented by green). This is due to the introduction of BMXBMI drastically reducing the overall indistinguishability of the tuples in the dataset. However, BPQ020, among other attributes in the dataset, has much less impact on reducing the tuple indistinguishability. Hence, the comparison between Figures \ref{csfplot} and \ref{pifplot} provides a clear indication of the intuition behind the $PIF$ value generation. As shown in Figure \ref{pifplot}, higher $PIF$ values indicate higher levels of distinguishability (or $PIF$ leak) compared to the other attributes.

The separate analysis of the $Q$ attributes (represented by the red bars in Figure \ref{pifplot}) provides a clearer understanding of their impact on $PIF$ values compared to when they are introduced to the $S$ attributes, as demonstrated in Figures \ref{csfplotqat} and \ref{pifplotqat}. It is clear that $PIF$ values of the attributes LBXTC and ALQ120Q  significantly increase when they are introduced to the $S$ attributes. 

Figure \ref{refcsfpifplotsq} shows the $CSF$ and $PIF$ dynamics of the refined set of $Q$ attributes. As depicted by the plots, OptimShare has identified that LBXTC and ALQ120Q should be removed from the set of $Q$ attributes as they leak too much information when released with no perturbation. Hence, LBXTC and ALQ120Q  are automatically considered as sensitive attributes and moved to the set of $S$ attributes. As shown in the plots (refer to Figure \ref{refcsfpifplotsq}), the refined $Q$ attributes show minimal data distinguishability, producing more homogeneity in the refined $Q$-dataset tuples. This result, in turn, supports the application of less perturbation on the $S$-dataset compared to the previous non-refined $Q$ attribute set.

Figure \ref{utilityeffectiveness} shows the utility and effectiveness variations of the 12 datasets produced for the twelve $\varepsilon,\delta$ combinations ($TN_{\varepsilon,\delta}=12$). As Figures \ref{utilityplot} and \ref{effectivenessplot} show, the utility and effectiveness of the dataset are almost similar. This is due to the corresponding datasets producing much lower normalized residual linkage leak ($P_N$) than the utility values. This also suggests that OptimShare effectively refined the $Q$ attribute, so the datasets can still maintain a lower residual linkage leak.

We compared DP-WGAN and OptimShare using the NHDS dataset, applying the configurations from Section \ref{OptimShareconfig} to generate privacy-preserving datasets. Upon comparing these with the original dataset, the mean and standard deviation of two $Q$-attributes (BPQ020 and RIAGENDR) were evaluated. DP-WGAN's perturbation of the entire dataset largely destroys individual statistics for these attributes, whereas OptimShare preserves them, yielding nearly identical utility to the original except for a small discrepancy, mostly due to replacing missing values. We also measured the Na\"\i ve Bayes classification performance across the three datasets using RMSE, precision, and recall. Interestingly, DP-WGAN outperforms the original dataset due to the beneficial effect of perturbation on the dataset's distribution. However, OptimShare still delivers good performance on all three metrics, benefiting from perturbation and the preservation of certain attributes. This indicates OptimShare's high utility and potential applicability in scenarios requiring accurate global statistics, such as tracking exact COVID-19 case distribution across postcodes, assuming postcode attributes meet OptimShare's indistinguishability requirements.

\begin{table}
\vspace{-0.5cm}
\centering
\caption{Comparison of results of OptimShare against direct perturbation by DP-WGAN. The following experiments were carried out under the same configurations explained in Section \ref{OptimShareconfig}. NB represents Na\"ive Bayes classification, and std. represents the standard deviation.}   
    \label{compareresults}
    \begin{small} 
        \resizebox{0.5\columnwidth}{!}{
\begin{tabular}{@{}lccc@{}}
\toprule
                                       & \textbf{Original} & \textbf{DP-WGAN} & \textbf{OptimShare} \\ \midrule
BPQ020\_mean   & 0.0117            & 6.72105          & 0.0079              \\
RIAGENDR\_mean & 1.30235           & 1.66805          & 1.35855             \\
BPQ020\_std.                           & 0.6667            & 0.97205          & 0.6667              \\
RIAGENDR\_std.                         & 0.45205           & 0.1451           & 0.47925             \\
NB\_RMISE                              & 0.4015            & 0.4125           & 0.4153              \\
NB\_precision                          & 0.760             & 0.780            & 0.811               \\
NB\_recall                             & 0.789             & 0.794            & 0.798               \\ \bottomrule
\end{tabular}
}
\end{small}
\vspace{-1cm}
\end{table}

\section{Related Works}
Literature shows a few attempts to utilize data perturbation to solve privacy issues in tabular data-sharing (non-interactive data sharing) for different application-specific scenarios. Two of the primary advantages of data perturbation against cryptographic protocols are efficiency and scalability. Examples of data perturbation techniques include additive perturbation, random rotation, geometric perturbation, randomized response, random projection, microaggregation, hybrid perturbation, data condensation, data wrapping, data rounding, and data swapping~\cite{chamikara2020efficient,dwork2014algorithmic,hasan2016effective}. However, the utilization of these perturbation techniques is often intended for one application (e.g., histogram analysis, deep learning), which restricts the utility of the corresponding perturbation approach to one dedicated task. Hence, the generalizability of a perturbation mechanism has not been of fundamental focus, and the practicality of these approaches for real-world applications has been a challenge. The literature does not show many mechanisms that have been developed to investigate the tradeoff between utility and privacy~\cite{xu2015privacy} related to tabular data perturbation, towards supporting practical utility (i.e., not restricting the utility to one application). Bertino et al.'s framework for evaluating privacy-preserving data mining algorithms is one of the few approaches developed to evaluate the balance between privacy and utility. However, their approach is more of a perturbation quality evaluation approach than an approach to improve the usability (practical utility) of data perturbation approaches~\cite{bertino2005framework}. FRAPP is another solution that provides matrix-theoretic framework-based solutions for random perturbation schemes~\cite{agrawal2005framework}. Thuraisingham et al. attempted to develop insights into balancing privacy and utility during privacy preservation~\cite{thuraisingham2017towards}. Although these approaches are insightful, they did not specifically answer the usability aspect of a perturbation mechanism in the real-world setting. Although a few other  framework-based solutions, such as PSI ($\Psi$~\cite{gaboardi2016psi}), investigate the generalizability of data sharing with high privacy, they often focus only on the interactive data sharing setting. Hence, it is essential to develop a unified framework-based solution to improve the practical utility of non-interactive privacy preservation mechanisms.

\section{Conclusion}
This paper introduces OptimShare, a unified framework-based solution for privacy-preserving tabular data sharing. Unlike existing methods that concentrate on one problem (e.g., histogram analysis), OptimShare caters to a wide range of use cases, resolving privacy and utility issues more effectively. OptimShare uniquely identifies the privacy requirements of a specific dataset through a novel approach called the Personal Information Factor (PIF) and allows a carefully selected limited set of raw attributes to be released, adhering to differential privacy principles. OptimShare achieves this by a rigorous iterative privacy enforcement mechanism, yielding a perfect balance between privacy and utility. This is verified by the empirical evidence produced by OptimShare. Lastly, we developed both web-based and stand-alone versions of OptimShare. In particular, the web-based system focuses more on security by isolating raw datasets according to roles, and the system enables scalability and CI/CD using Docker containers.

Despite OptimShare's effective approach to handling controlled partially perturbed non-interactive data sharing (CPNDS),  CPNDS still introduces new challenges around maintaining a proper balance between utility and privacy. This arises largely from the complexity of input data that presents unlimited potential scenarios, suggesting avenues for future work.  We continuously examine these dynamics as part of our development and strive for ongoing OptimShare improvements.

\section{Acknowledgment}
The work has been supported by the Cyber Security Research Centre Limited whose activities are partially funded by the Australian Government’s Cooperative Research Centres Programme.

\section{Appendices}

\subsection{proofs}
\subsubsection{Proof 1}
\label{prooflinkability}
\begin{proof}
Consider $D$ as an original dataset with $n$ tuples and $m$ attributes. Define $S$ and $Q$ as sets of sensitive and non-sensitive attributes in $D$ respectively. Assume the adversary possesses complete knowledge of $Q$ in perturbed dataset, $D^p$. 

We define record linkability as follows. Consider $R$ as the collection of all records in $D$ and $D^p$. If $q^\alpha = q^\beta$ for some $q \in Q$ and $\alpha, \beta \in R$, then $(q^\alpha, s^\alpha)$ and $(q^\beta, s^\beta)$ are part of the same similarity group, $SG$. Compute the cosine similarity, $CS^i_k$, between original and perturbed $S$ attributes of each record $r_i$ in $SG_k$. A record is linkable if $CS^i_k \leq CS^j_k$ for all $i \in R_{SG_k}$, for some $j \in R_{SG_k}$. Denote linkable records set as $L$.

$\varepsilon$-differential privacy is satisfied if for any datasets $D_1$ and $D_2$ differing by at most one record, and any outcome $o$ of a randomized algorithm $M$, the following inequality holds:

\begin{equation}
\frac{P[M(D_1) = o]}{P[M(D_2) = o]} \leq e^\varepsilon
\label{dpmodel}
\end{equation}

Take $D_1$ as the original dataset and $D_2$ as the dataset identical to $D_1$ but with modified sensitive attributes in one record. Then, we can apply $\varepsilon$-differential privacy, showing the adversary's successful record linkage probability is minimal.

Calculate the probabilities in the inequality's numerator and denominator. The numerator's probability is the chance that $D^p$ contains a record $(q^\alpha, s^\alpha)$ in the same $SG$ as $(q^\beta, s^\beta)$, and $(q^\alpha, s^\alpha)$ is linkable. This is:
\begin{equation}
P[M(D_1) = o] = P[(q^\alpha, s^\alpha) \in SG \ \land \ CS^i_k \leq CS^j_k \ \forall \ j \in R_{SG_k}]
\end{equation}

For the denominator, the probability is the chance that $D^p$ contains a record $(q^\alpha, s^\beta)$ in the same $SG$ as $(q^\beta, s^\beta)$, and $(q^\alpha, s^\beta)$ is linkable:
\begin{equation}
P[M(D_2) = o] = P[(q^\alpha, s^\beta) \in SG \ \land \ CS^i_k \leq CS^j_k \ \forall \ j \in R_{SG_k}]
\end{equation}

Substituting into Equation \ref{dpmodel}, we get:

\begin{equation}
\frac{P[(q^\alpha, s^\alpha) \in SG \ \land \ CS^i_k \leq CS^j_k \ \forall \ j \in R_{SG_k}]}{P[(q^\alpha, s^\beta) \in SG \ \land \ CS^i_k \leq CS^j_k \ \forall \ j \in R_{SG_k}]} \leq e^\varepsilon
\end{equation}

This suggests the adversary's successful record linking probability is limited, fulfilling the $\varepsilon$-differential privacy requirement.

\end{proof}

\subsubsection{Proof 2}
\label{optimsharedpproof}
\begin{proof}

The proof of Theorem \ref{theroemdp} requires demonstrating the numerator and denominator of the Theorem's Equation are small, indicating the probability of a record in a similarity group being linkable is minimal. This necessitates verifying that the perturbations on $D^p$'s sensitive attributes suffice to deter successful record linking by an adversary.

This is feasible by ensuring the cosine similarity between the original and perturbed sensitive attributes of all $D^p$ records is minimal. Lower cosine similarity complicates record linking for the adversary as it dictates the record's linkability probability. Compliance with the privacy budget demands a negligible change in a specific outcome's probability when a record is added or deleted, which is achievable by applying DP noise to sensitive attributes during perturbation.

The sufficiently small cosine similarity between original and perturbed attributes can be upper-bounded using record linkability (Definition \ref{deflink}), computing the cosine similarity for each dataset record. Complying with the privacy budget involves bounding the change in a specific outcome's probability upon record addition or deletion.

Considering two records, $(q_1, s_1)$ and $(q_2, s_1')$, which have identical quasi-identifiers, and sensitive attributes $s_1$ and $s_1'$, (where $s_1'$ is the perturbed version of $s_1$, generated using an $(\varepsilon, \delta)$-differentially private generator), we can compute the cosine similarity of original and perturbed sensitive attributes, showing the insignificant change in a specific outcome's probability with record addition or deletion.

The cosine similarity between $s_1$ and $s_1'$ is calculated as:

\begin{equation}
CS = \frac{s_1 \cdot s_1'}{|s_1| |s_1'|}
\end{equation}

We can use the Cauchy-Schwarz inequality~\cite{bhatia1995cauchy} to show that: 

\begin{equation}
s_1 \cdot s_1' \leq |s_1| |s_1'|
\end{equation}

Given the constraints set by $\frac{\varepsilon L}{T_{\varepsilon}}$ (where $L$ represents the set of linkable records), we can establish an upper bound for $|s_1'|$ to ensure that the cosine similarity is small.

For $\frac{\varepsilon L}{T_{\varepsilon}} \leq 1$, we can ensure that the added noise is within the acceptable range defined by $\varepsilon$. This limits the denominator of the cosine similarity expression to a value that's consistent with the privacy budget, $\varepsilon$.

Therefore, the cosine similarity between the original and perturbed sensitive attributes is upper-bounded by a value that complies with the privacy budget $\varepsilon$, which confirms that the OptimShare framework satisfies $\varepsilon$-differential privacy.

\end{proof}

\subsection{Different Interface Views of the OptimShare Live Tool}
\label{experimentalplots}
\subsubsection{OptimShare web-based and stand-alone live tool}
\begin{figure}[H]
\vspace{-1cm}
    \subfloat[View of a dataset on the server]{
        \includegraphics[width=0.22\textwidth, trim=0cm 0cm 0cm 0cm]{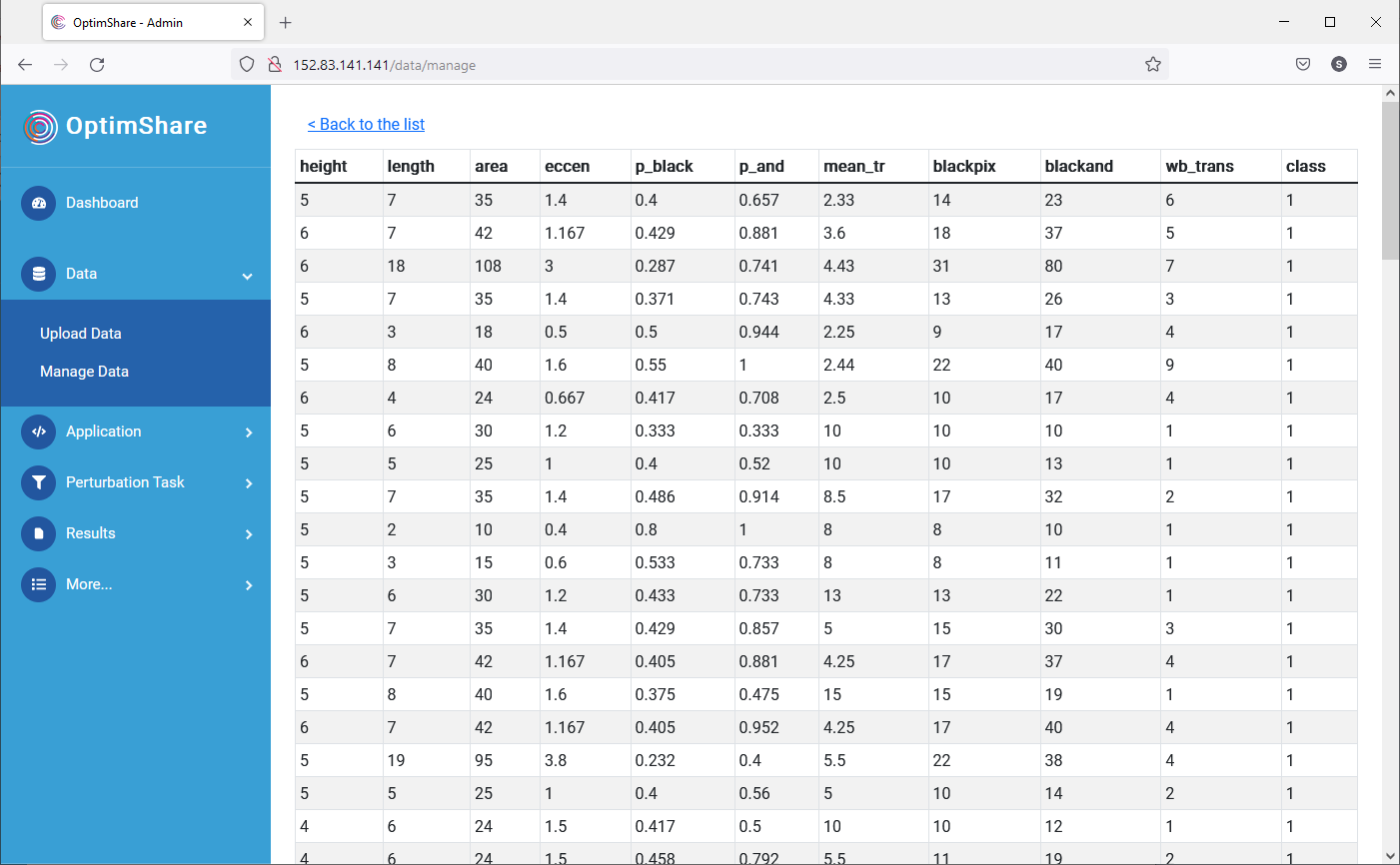}
        \label{webviewdata}}
    \hfill
    \subfloat[Details of an privacy-preserving algorithm]{
        \includegraphics[width=0.22\textwidth, trim=0cm 0cm 0cm 0cm]{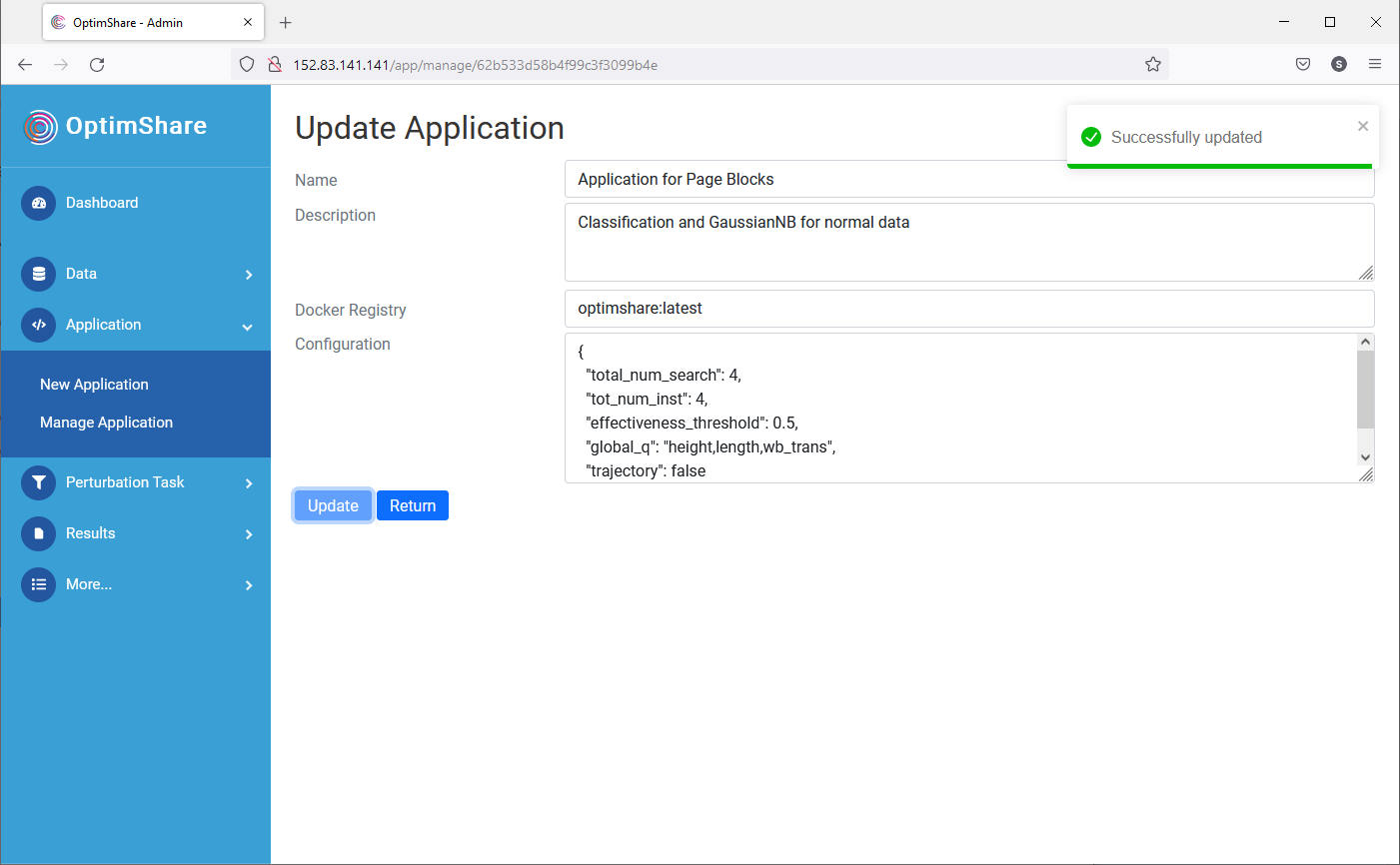}
        \label{webapplicationdetail}}
    \hfill
    \subfloat[Detail of a perturbation task]{
        \includegraphics[width=0.22\textwidth, trim=0cm 0cm 0cm 0cm]{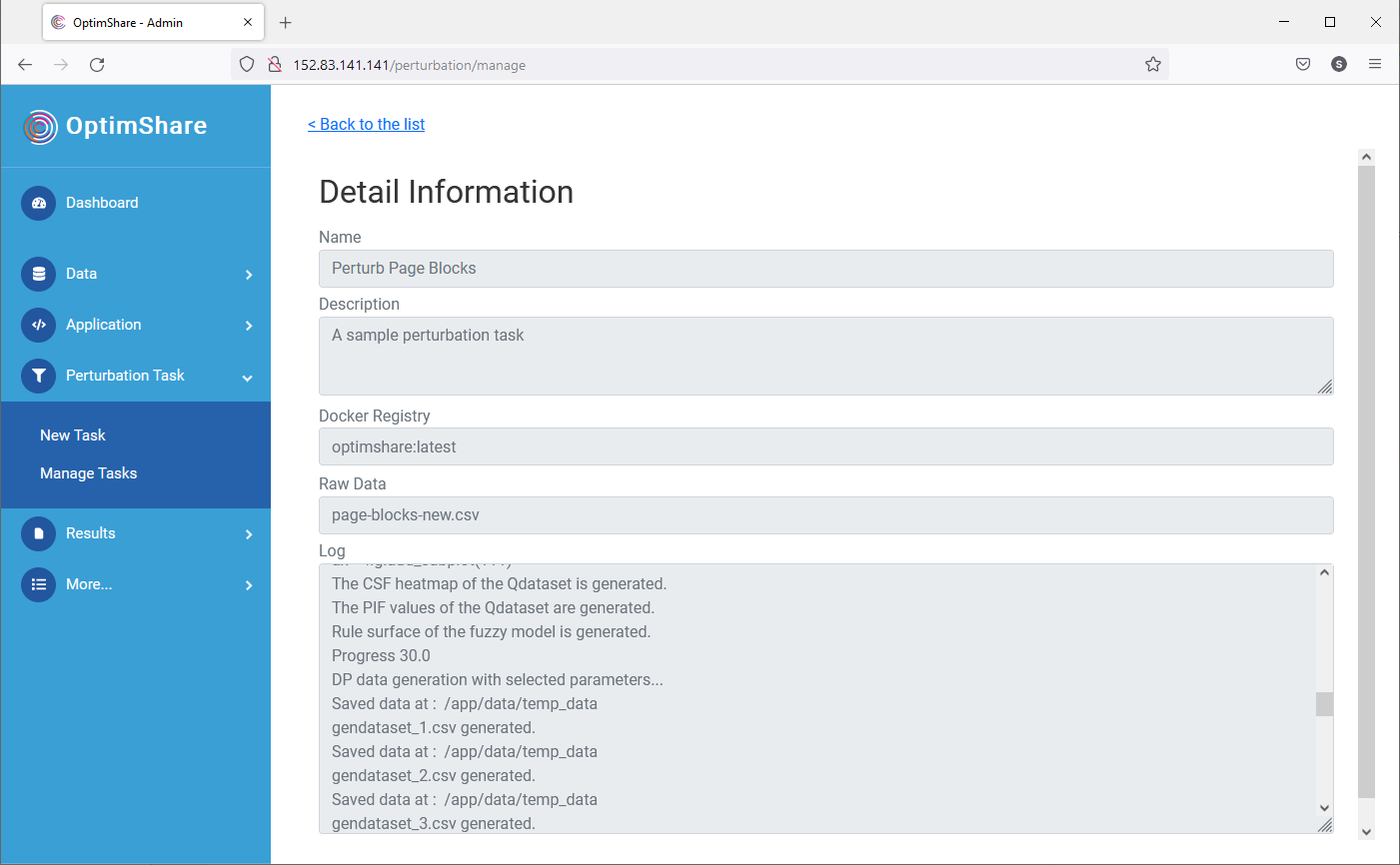}
        \label{webrunning}}
    \hfill
    \subfloat[Published dataset to data users]{
        \includegraphics[width=0.22\textwidth, trim=0cm 0cm 0cm 0cm]{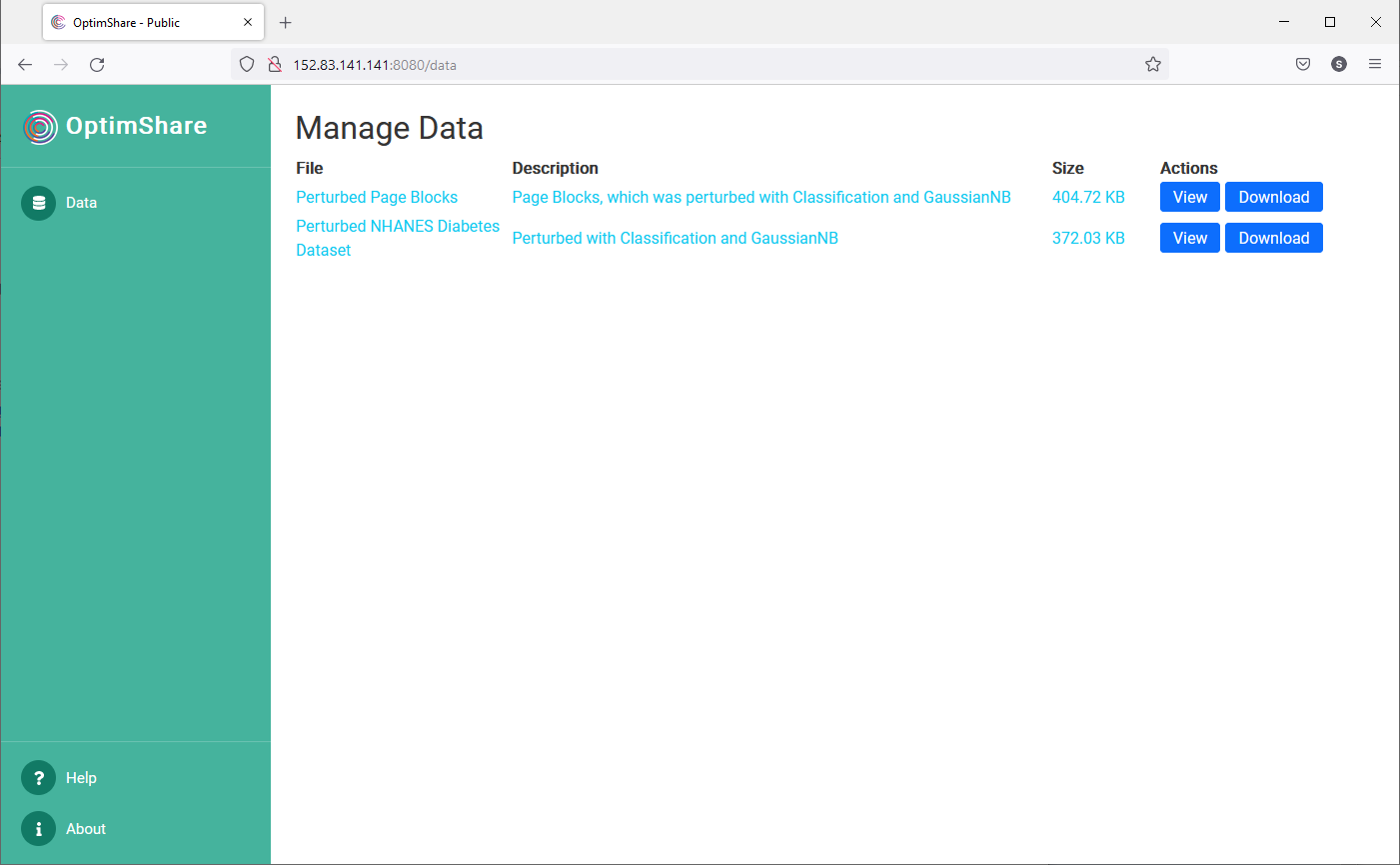}
        \label{webpublic}}
    \caption{Screenshots of the server-based OptimShare and the web server for Data Users.}
    \label{serverversion}
    \medskip
    \small
    \vspace{4cm}
\end{figure}

\begin{figure}[H]
    \subfloat[A loaded dataset in the Stand-alone version]{
        \includegraphics[width=0.15\textwidth, trim=0cm 0cm 0cm 0cm]{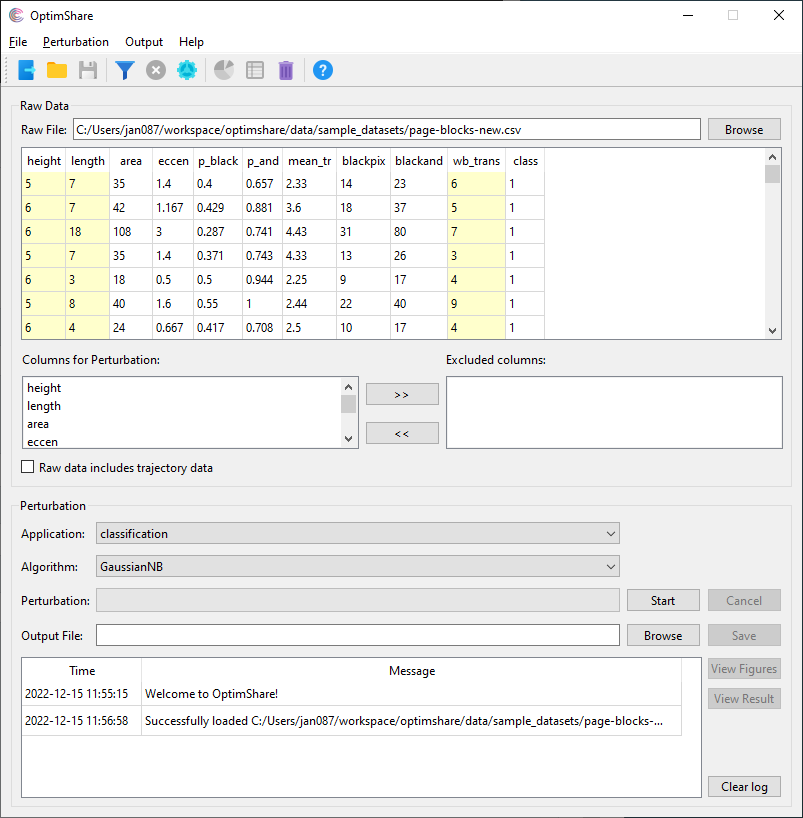}
        \label{aloneviewdata}}
    \hfill
    \subfloat[Applying a privacy-preserving algorithm on the dataset]{
        \includegraphics[width=0.15\textwidth, trim=0cm 0cm 0cm 0cm]{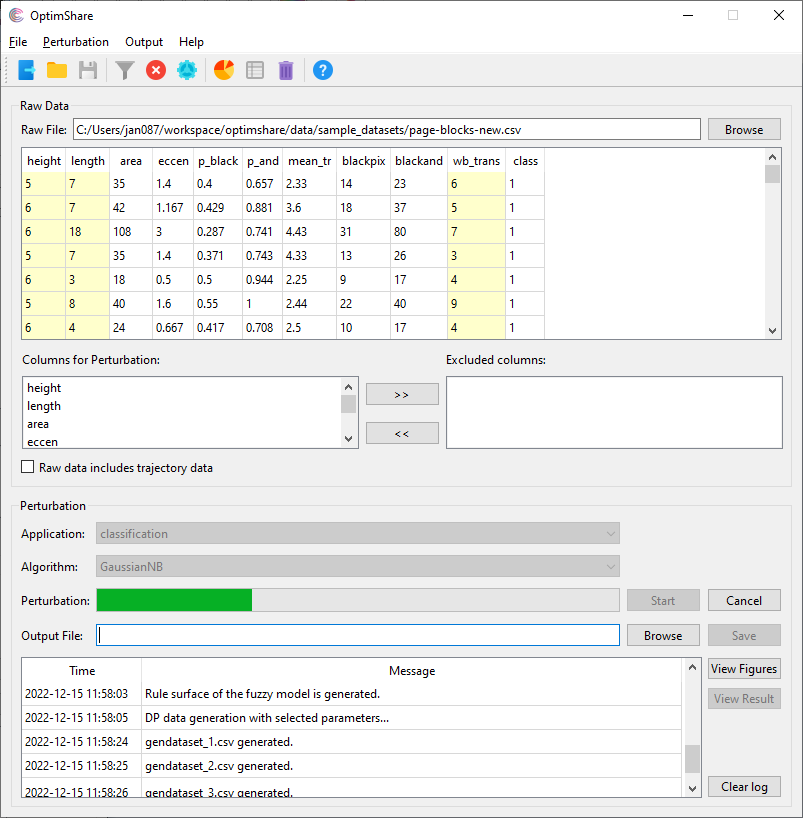}
        \label{alonerunning}}
    \hfill
    \subfloat[Intermediate figures]{
        \includegraphics[width=0.15\textwidth, trim=0cm 0cm 0cm 0cm]{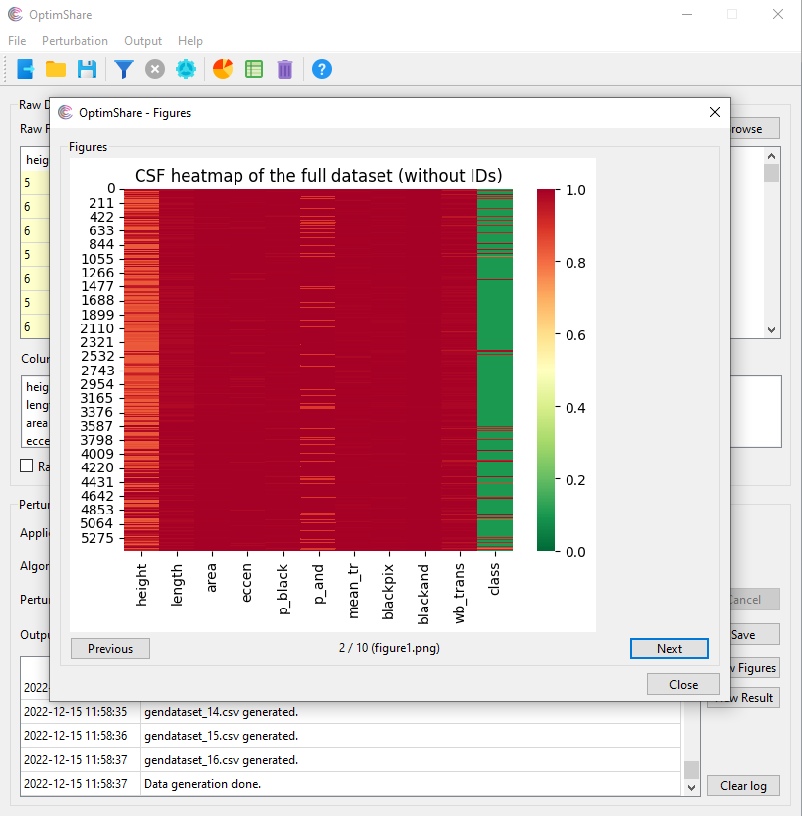}
        \label{alonefigures}}
    \hfill  
    \subfloat[Extended configurations]{
        \includegraphics[width=0.18\textwidth, trim=0cm 0cm 0cm 0cm]{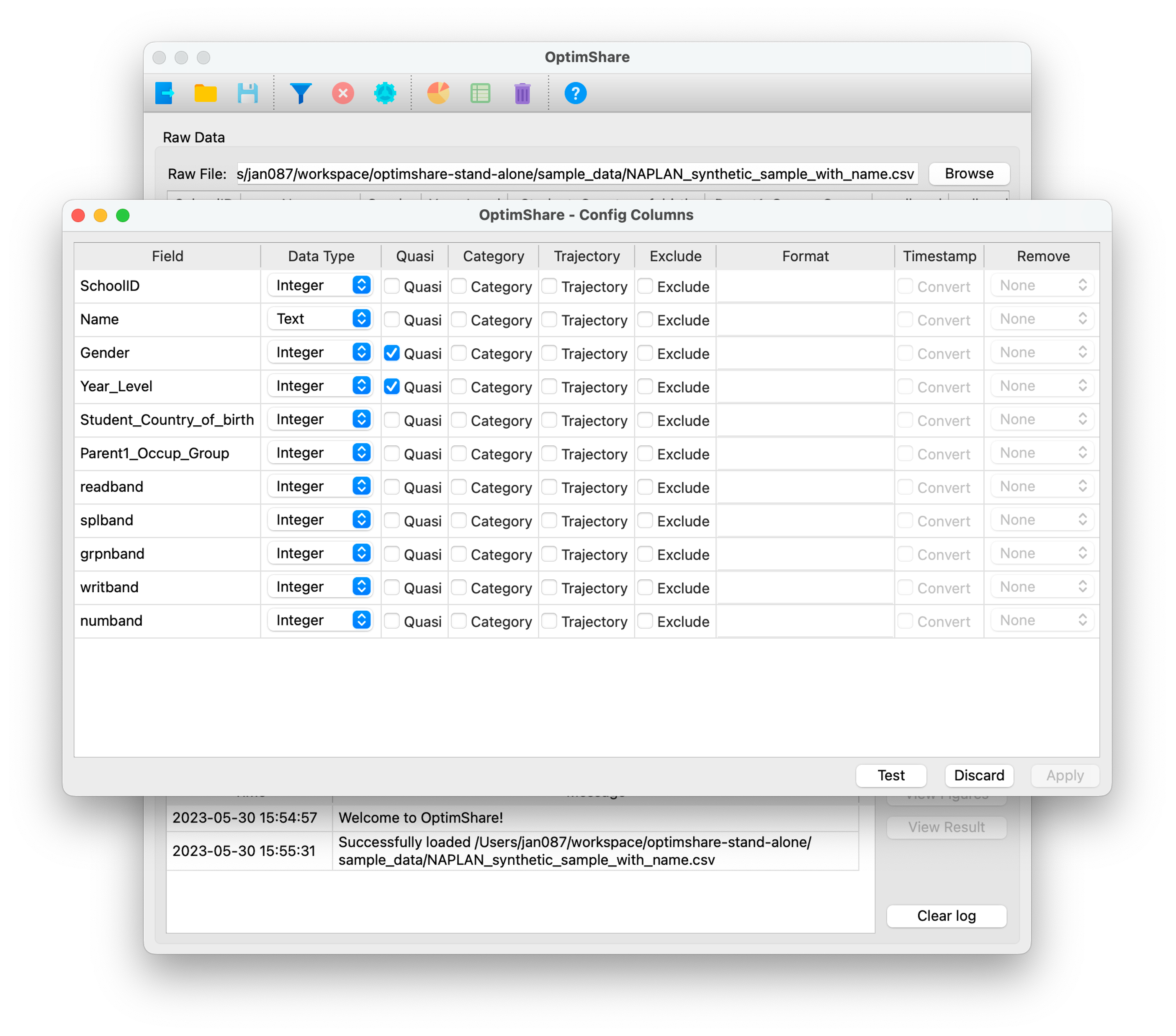}
        \label{configurations}}
    \hfill
    \subfloat[A perturbed dataset]{
        \includegraphics[width=0.15\textwidth, trim=0cm 0cm 0cm 0cm]{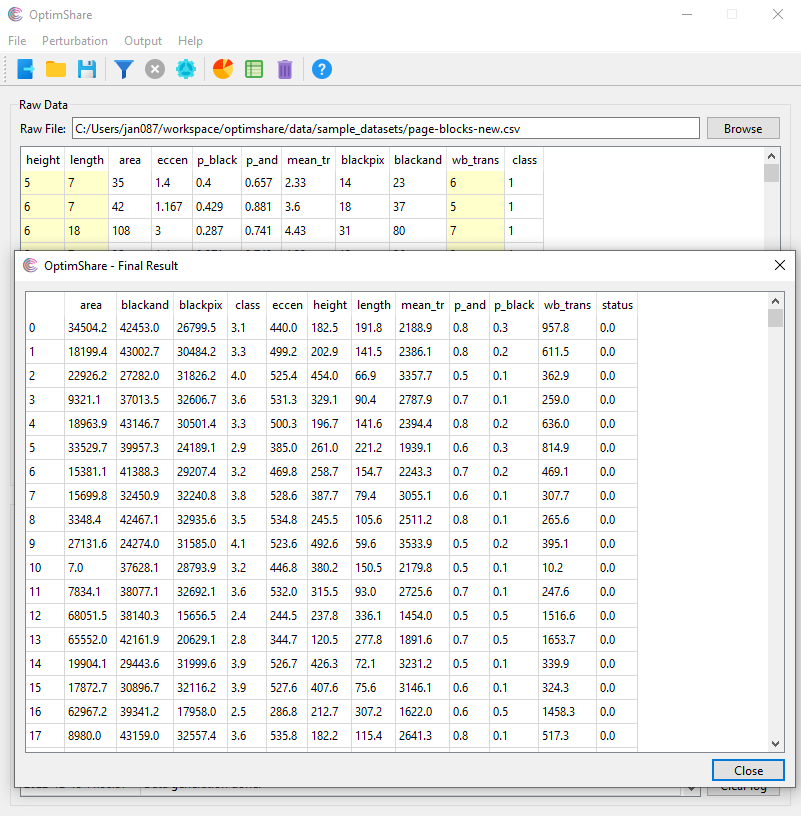}
        \label{aloneperturbed}}
    \hfill
    \caption{Screenshots of the OptimShare stand-alone version.}
    \label{standalone}
    \medskip
    \small
\end{figure}

Figure \ref{serverversion} and Figure \ref{standalone} show the  screenshots of the two versions (web-based and stand-alone) of the OptimShare live tool.


\end{document}